\documentclass[twocolumn]{aastex61}

\usepackage{natbib}
\usepackage{graphicx}
\usepackage{float}
\usepackage{amsmath,amstext,amssymb}
\usepackage{hyperref}

\shorttitle{Narrow-Line Region of QSO2s}
\shortauthors{Storchi-Bergmann et al.}

\begin{document}

\title[Narrow Line Region of nearby QSO2s]{Bipolar ionization cones in the Extended Narrow-Line Region of nearby QSO2s}

\author{T. ~Storchi-Bergmann}
\affil{Departamento de Astronomia, Universidade Federal do Rio Grande do Sul, IF, CP 15051, 91501-970 Porto Alegre, RS, Brazil}
\affil{Harvard-Smithsonian Center for Astrophysics, 60 Garden St., Cambridge, MA 02138, USA}

\author{B. Dall'Agnol de Oliveira}
\affil{Departamento de Astronomia, Universidade Federal do Rio Grande do Sul, IF, CP 15051, 91501-970 Porto Alegre, RS, Brazil}

\author{L. F. Longo Micchi}
\affil{Institute for Astrophysics and Computational Sciences, Department of Physics, The Catholic University of America, Washington, \\ DC 20064, USA}

\author{H. R. Schmitt}
\affil{Naval Research Laboratory, Washington, DC 20375, USA 0000-0001-7376-8481}

\author{T. C. Fischer}
\affil{Goddard Space Flight Center, 8800 Greenbelt Rd, Greenbelt, MD 20771, USA}

\author{S. Kraemer}
\affil{Institute for Astrophysics and Computational Sciences, Department of Physics, The Catholic University of America, Washington, \\ DC 20064, USA}

\author{M. Crenshaw}
\affil{Department of Physics and Astronomy, Georgia State University, Astronomy Offices, 25 Park Place, Suite 600, Atlanta, GA 30303, \\USA}

\author{P. Maksym}
\affil{Harvard-Smithsonian Center for Astrophysics, 60 Garden St., Cambridge, MA 02138, USA}

\author{M. Elvis}
\affil{Harvard-Smithsonian Center for Astrophysics, 60 Garden St., Cambridge, MA 02138, USA}

\author{G. Fabbiano}
\affil{Harvard-Smithsonian Center for Astrophysics, 60 Garden St., Cambridge, MA 02138, USA}

\author{L. Colina}
\affil{Centro de Astrobiología (CAB, CSIC-INTA), Carretera de Ajalvir, 28850 Torrejón de Ardoz, Madrid, Spain}

\correspondingauthor{T. ~Storchi-Bergmann}
\email{thaisa@ufrgs.br}

\begin{abstract}

We have used narrow-band [OIII]$\lambda\lambda$4959,5007 and H$\alpha$+[NII]$\lambda\lambda6548,84$ Hubble Space Telescope (HST) images of 9 luminous (L[OIII]$>10^{42}$erg\,s$^{-1}$) type 2 QSOs with redshifts $0.1<z<0.5$ in order to constrain the geometry of their Extended Narrow-Line Regions (ENLR), as recent ground-based studies suggest these regions become more spherical at high luminosities 
due to destruction of the torus.  We find instead elongated ENLRs reaching 4 to 19 kpc from the nucleus and bipolar ionization cones in [OIII]/(H$\alpha$+[NII]) excitation maps indicating that the torus survives these luminosities, allowing the escape of $\approx$\,10 times higher ionizing photon rates along the ionization axis than perpendicularly to it. The exceptional HST angular resolution was key to our success in arriving at these conclusions. Combining our measurements with previous ones based on similar HST data, we have revisited the relation between the ENLR radius R$_{maj}$ and L[OIII] over the range $39<$log(L[OIII])$<43.5$ (L in erg\,s$^{-1}$): log(R$_{maj}) = (0.51\pm0.03)$\,log(L[OIII])$-18.12\pm0.98$. The radius of the ENLR keeps increasing with L[OIII] in our data, implying that the ENLR can extend to distances beyond the limit of the galaxy if gas is present there -- e.g. from AGN outflows or interactions, seen in 6 objects of our sample. We attribute the flattening previously seen in this relation to the fact that the ENLR is matter-bounded,
meaning that ionizing photons usually escape to the intergalactic medium in luminous AGN. Estimated ionized gas masses of the ENLRs range from 0.3 to $2\times10^8$\,M$_{\odot}$, and estimated powers for associated outflows range from $<0.1\%$ to a few percent of the QSO luminosity.


\end{abstract}


\section{Introduction}

The physical processes that couple the growth of supermassive black holes (SMBH) to their host galaxies 
occur in the vicinity of the galaxy nucleus ($\approx$\,inner kpc) \citep{hop10} when it becomes active due to mass accretion to the SMBH \citep{fer05,kor13}. The radiation emitted by Active Galactic Nuclei (AGN) works as a flashlight that illuminates and ionizes the gas in the vicinity of the nucleus, forming the Narrow Line Region (NLR). AGN-driven winds \citep{elv00,cio10,sto10,barbosa14} interact with the gas and produce outflows reaching velocities of hundreds of km\,s$^{-1}$ \citep{fis13,sto10}. Relativistic jets emanating from the AGN also interact with the gas of the NLR \citep{wan09,couto17}. Both types of outflow produce feedback, which has been an important ingredient in galaxy evolution models to avoid producing over-massive galaxies \citep{fab12}. Inflows have also been observed in this region \citep{rif08,rif13,cre10,schnorr14a}. The importance of the NLR stems from the fact that it is spatially resolved, as it extends from hundreds to thousands of parsecs from the nucleus and exhibits strong line emission. These properties of the NLR allow the observation  of the interaction between the AGN and the circumnuclear gas in the galaxy:  (1) via the observation of the NLR geometry and excitation properties that constrain the AGN structure and nature of the ionizing source; (2) via the gas kinematics that maps the processes of the AGN feeding and feedback.

Over the last 30 years, narrow-band imaging of the Narrow Line Region (NLR) became an important tool in the study of their ionization mechanism, morphology and implied geometry for the AGN. Starting with ground based observations of Seyfert (Sy) galaxies \citep[e.g.][]{pog88,han88,wil94}, emission-line images of [OIII]$\lambda$5007 and H$\alpha$+[NII] showed that the NLRs of several Sy galaxies have biconical shapes, with the apex at the nucleus. These observations were of key importance in validating the Unified Model of AGN \citep{ant93}, in which both Sy types have a SMBH fed by an accretion disk, surrounded by a torus of molecular gas and dust that collimates the ionizing radiation leading to the observed biconical shape. However, although these ground-based narrow-band images of Sy galaxies have allowed the detection of shadowing of the nuclear ionizing radiation, they could resolve only the nearest and brightest sources. For fainter and more distant sources, the NLR angular diameters decrease below the resolution limit of ground-based telescopes ($\sim$\,1$^{\prime\prime}$).
The use of the Hubble Space Telescope (hereafter HST) to image the NLR was crucial for the further development of the field \citep[e.g.][]{wil93,cap96,fal98,fer00,sch03a}. The high spatial resolution of these observations allowed better constraints on the Unified Model, and the study the NLR structure in detail.  Some  results from these studies include the alignment of the radio jet and NLR, which indicate that the torus and accretion disk are connected; that the radio jet can significantly influence the NLR emission; and that Sy 1 NLRs are statistically more circular and concentrated than Seyfert 2's, which can be attributed to foreshortening in the former \citep{sch03b}.

Another important result from these studies -- in particular of the snapshot survey of \citet{sch03a,sch03b} -- was the detection of a relation between the NLR radius R$_{NLR}$ and its luminosity L([OIII]). For a sample of 60 AGN with L([OIII])$<\,10^{42}$\,erg\,s$^{-1}$, these authors found the relation R$_{NLR}\propto$\,L[OIII]$^{0.33}$. This was  interpreted as due to the fact that the ionization parameter is not constant along the NLR, and that most of the [OIII] emission comes from a low density region. A constant ionization parameter had been suggested by previous observations of the NLR of a sample of QSO's that implied a slope of $\sim$0.5 for the relation above \citep{ben02}, but which was later concluded to lead to predicted NLR sizes much larger than observed \citep{net04}. 



Recent results obtained via Integral Field Spectroscopy \citep[e.g.][]{liu13} have suggested that at higher redshifts and luminosities, the geometry of the AGN outflows may change, becoming less collimated and more spherical, due to the higher AGN luminosity and higher power of the possibly associated outflow, that would clear a larger area of the dusty toroidal region surrounding the accretion disk and broad-line region (BLR) \citep{elitzur14}. But, as these studies have been done using ground-based telescopes, for $z\ge0.1$, the limited angular resolution and smearing by the seeing precludes firm constraints on the NLR`(or ENLR) morphology. In order to provide such constraints at higher redshifts and AGN luminosities, we have obtained narrow-band [OIII]$\lambda$5007 and H$\alpha$+[NII] images with the Hubble Space Telescope (HST) of a flux-limited sample of 9 luminous type 2 AGN (or QSOs, due to their high luminosites), with redshifts in the range $0.1\le z \le 0.5$. We use these observations to obtain excitation maps, through the ratio [OIII]/(H$\alpha$+[NII]), ionized gas masses from the H$\alpha$ luminosities and extend the relation between the ENLR radius R$_{maj}$ and L[OIII], from 39$<$log(L[OIII])$<$43.5 (L in erg\,s$^{-1}$).

This paper is organized as follows: in Sec.\ref{sec:Sample} we present the sample, in Sec. \ref{sec:observations} we discuss the observations, data  reduction and measurements, in Sec. \ref{sec:results} we present the images and ionization maps, as well as total gas masses and spatial profiles of the ionized gas mass surface density. In Sec. {discussion}  we discuss our results and in Sec. \ref{sec:conclusions}  we present our conclusions.

\section {Sample}\label{sec:Sample}

Our sample was selected from the catalogue of QSO2  galaxies of \citet{rey08}, that comprises 887 galaxies with $0.04<z<0.8$, thus just beyond the redshift range of the HST imaging survey of \citet{sch03a}. The selection was done according to the following criteria: (1) the galaxies should have [OIII] fluxes larger than 1.5$\times$\,10$^{-14}$ ergs\,cm$^{-2}$\,s$^{-1}$ in order to assure enough signal-to-noise ratio in the narrow-band images; (2) they should have  redshifts in the range $0.1<z\,\le\,0.5$, thus allowing to probe more luminous AGNs than in the previous study of  \citet{sch03a}; the luminous sources at z\,$\approx$\,0.4 are predicted to  have NLR sizes of $\sim$\,3\,kpc, that correspond to angular sizes of about 0.6 arcsec, thus resolvable down to hundred of parsec scales at the host galaxy with HST;
(3) having L[OIII] values larger than $\mathrm{log[L[OIII]=42.5}$ (in $\mathrm{erg\,s^{-1}}$). 

The list of the sample galaxies, together with their redshifts, 
angular scales (obtained from the angular distances) and luminosity distances ($D_L$) are shown in Table \ref{tab:sample}. The listed redshifts and the other derived quantities were obtained from the NASA/IPAC Extragalactic Database (hereafter NED). The angular and luminosity distances were corrected to the Cosmic Microwave Background Radiation reference frame.
The uncertainties shown in the quantities listed in the Table\,\ref{tab:sample} correspond to the maximum errors resulting from the uncertainties in $H_0$ and z. 
The spatial scales range from 1.9 to 5.6\,kpc\,arcsec$^{-1}$, allowing a spatial resolution at the galaxies of 95\,pc to 280\,pc at the galaxies with HST.

We show in Fig.\,\ref{fig:specpanel}, the SDSS spectra of the sample galaxies together with curves showing the narrow filters passbands (see next section) used to obtain the line emission images centered on  [OIII]$\lambda5007$ and H$\alpha$+[NII], as well as the broad-band filter passbands used to obtain images in the continuum. For a few galaxies, the SDSS spectra did not cover the H$\alpha$+[NII] emission lines, as their redshifts put these lines beyond the upper wavelength limit of the SDSS spectra. The passbands were obtained with the Python package Astrolib PySynphot, which produces a synthetic throughput of the filter.

\begin{deluxetable}{ccccc}
\tablecolumns{5} 
\tablewidth{0pt} 
\tablecaption{Sample properties \label{tab:sample}}
\tablehead{
ID  & Name           & z     & Scale           & $D_L$ \\
(1)            & (2)   & (3)             & (4) &  (5)}
\startdata
1& J082313.50+313203.7 & 0.433 & 5.46$\pm$0.25   & 2310$\pm$110 \\
2& J084135.04+010156.3 & 0.111 & 1.954$\pm$0.093 & 498$\pm$24   \\
3& J085829.58+441734.7 & 0.454 & 5.61$\pm$0.25   & 2450$\pm$120 \\
4& J094521.34+173753.3 & 0.128 & 2.22$\pm$0.11   & 583$\pm$26   \\
5& J110952.82+423315.6 & 0.262 & 3.91$\pm$0.18   & 1284$\pm$58  \\
6& J113710.77+573158.7 & 0.395 & 5.16$\pm$0.24   & 2076$\pm$95  \\
7& J123006.79+394319.3 & 0.407 & 5.26$\pm$0.23   & 2149$\pm$93  \\
8& J135251.21+654113.2 & 0.206 & 3.26$\pm$0.16   & 980$\pm$47   \\
9& J155019.95+243238.7 & 0.143 & 2.42$\pm$0.12   & 653$\pm$32   \\
\enddata 
\tablecomments{
 (1) galaxy identification number used througout the paper;
 (2) galaxy identification (without \textit{SDSS} prefix); 
 (3) redshift; 
 (4) scale (in kpc/$^\prime\prime$);
 (5) luminosity distance (in Mpc).}
\end{deluxetable}

\section{Observations and Data Reduction}
\label{sec:observations}

The sample galaxies were observed with the HST Advanced Camera for Surveys (ACS) using linear ramp filters centered on the redshifted [OIII]$\lambda$5007 and H$\alpha+$[NII]$\lambda$6548,84 emission lines, as well as a medium or wide band filter centered in the continuum between these two sets of lines, used to subtract the continuum contribution underneath the emission lines. The log of the observations, executed under program GO-13741 (P.I. Storchi-Bergmann) is shown in Table\,\ref{tab:observation}.

\begin{figure*}
\includegraphics[width=.93\linewidth,angle=-0]{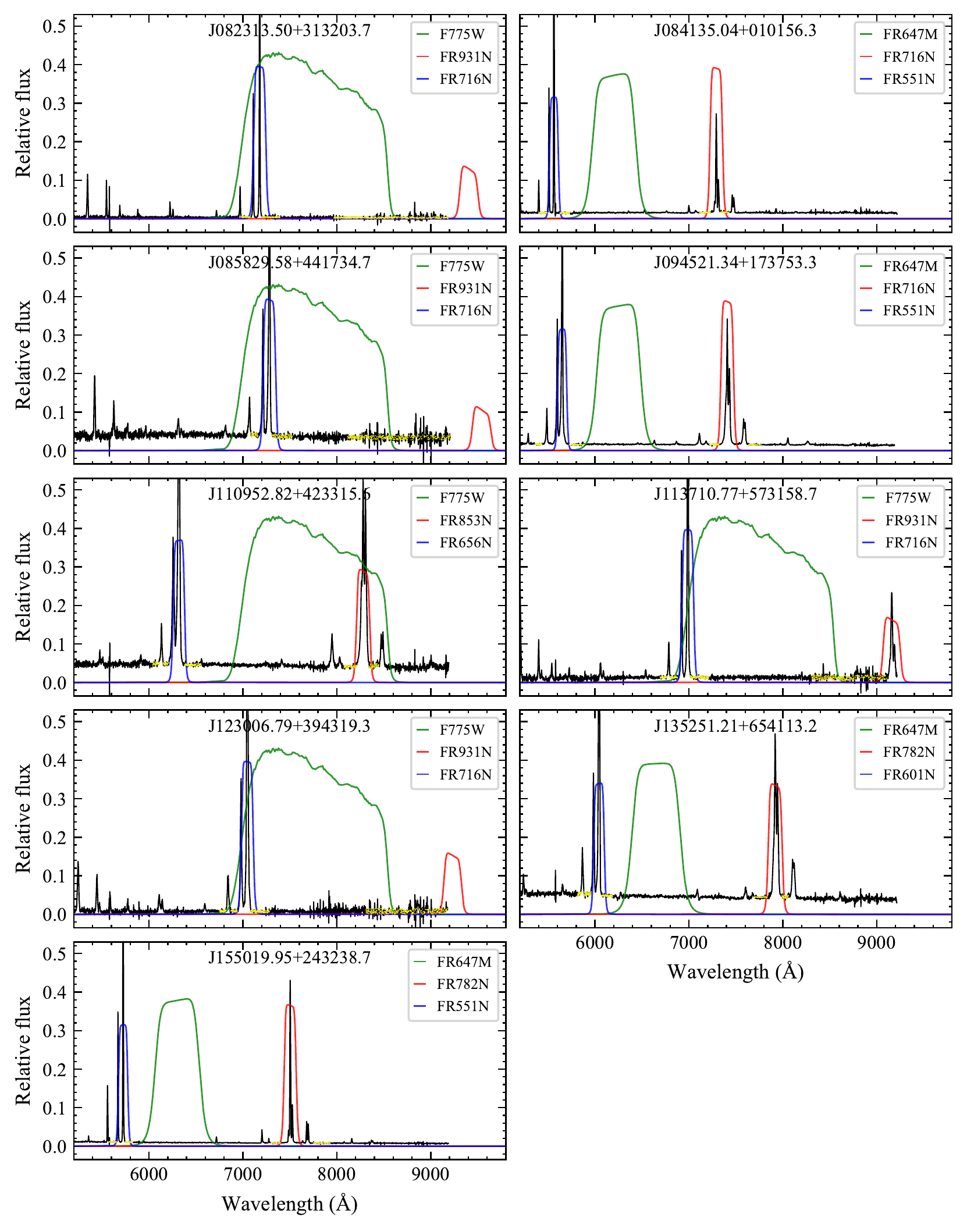}
\caption{SDSS spectra of the sample galaxies (normalized by the maximum flux). The colored curves illustrate the filter passbands used to obtain the corresponding HST images: [OIII] (blue curves), H$\alpha$+[NII] (red curves) and continuum (green curves). The yellow dots are the data used in the continuum fit for the calculation of S$_{C,[OIII]}$ and S$_{C,[NII]{+}H\alpha}$.}
\label{fig:specpanel}
\end{figure*}

The filter passbands are illustrated in Fig.\,\ref{fig:specpanel}, together with the SDSS spectra of each galaxy. It can be seen that, for all sources, the [OIII] filter covers well the 5007\AA\ emission line but only partially the 4959\AA\, line.  

Data reduction was performed using tasks in IRAF (Image Reduction and Analysis Facility). First, the set of images in each line were aligned with \textit{imalign}  and combined using  \textit{crrej}, which allows cosmic rays removal, and subsequently divided by the exposure time (header keyword \textit{EXPTIME}). Then, the image was multiplied by the header keyword \textit{PHOTFLAM}, to transform its units to $\mathrm{erg\,cm^{-2}\,s^{-1}\,\text{\AA}^{-1}}$. In order to allow for a precise continuum subtraction from the emission line images, the line images were aligned to the continuum images. In order to do this, the emission line images were rotated using the \textit{rotate} task, matching the continuum \textit{ORIENTAT} header keyword, and subsequently aligned with the continuum image with the task \textit{imalign}. After this, the sky contribution -- obtained as the center of a gaussian fit to the flux histogram distribution of a region devoid of galaxy contribution -- was subtracted from the image. For each image pixel, we assume a flux uncertainty equal to 3 times the standard deviation of this gaussian ($\sigma_{sky}$).

As can be observed in Figure\,\ref{fig:specpanel}, in a number of cases, the wide/median band filter used to obtain the continuum images includes also some contribution of emission lines. In order to correct for this contamination, we have used the SDSS spectra to calibrate our continuum images, by matching the continuum fluxes in the images to that of the SDSS spectrum  (that corresponds to a circular aperture with  diameter of 3 arcsec). This was done as follows. We fitted the SDSS continuum under the emission line [OIII]$\lambda4959{+}\lambda5007$, keeping only data differing by less than 2$\sigma$ of the average in the region. We then obtained the corresponding value from the HST continuum image ($C_{circ,[OIII]}$), by integrating the flux within a circular aperture with diameter of 3$''$ to match the SDSS aperture. The ratio S$_{C,[OIII]}$=C$_{SDSS,[OIII]}$/C$_{circ,[OIII]}$ gives the value used to scale the continuum image before subtracting it from the [OIII]$\lambda4959{+}\lambda5007$ image. The same procedure was performed to obtain S$_{C,([NII]{+}H\alpha)}$. In the cases for which these lines were not present in the spectra, we fitted a large region of the continuum in the red end of the spectrum (close to the missing lines). These values are presented in Table\,\ref{tab:reduction}, where the uncertainties in the SDSS continuum values were adopted as the standard deviation of the data used in the fit. These data are highlighted as yellow dots in Figure\,\ref{fig:specpanel}.

Table\,\ref{tab:reduction} shows that the smallest scale factors used to correct the continuum images for the emission-line contribution -- which correspond  to the largest corrections due to emission-line contamination -- were obtained for targets 7, 1, 6 and 3 (from the largest to the lowest corrections). In these cases, the continuum filter includes an important contribution from the [OIII] emission lines, ranging from 60\% to 20\% of the total flux.

As mentioned above, the [OIII] filter did not cover entirely the $\lambda$4959\AA\ emission line. In order to obtain [OIII] images including the total flux of both $\lambda$4959,$\lambda$5007 emission lines, we proceeded as follows. First, we subtracted the fitted continuum from each SDSS spectrum. Then, we multiplied the result by the respective [OIII] filter throughput and integrated in order to obtain the flux F[OIII$]_1$ that would be obtained using the [OIII] filter. Next, we modified the filter throughput, stretching the profile from the center by 50\AA\ toward smaller wavelengths, in  order to cover both emission lines. Now, the new integrated flux (F[OIII]$_2$) corresponds to what would be expected if the filter covered both emission lines. The ratio r$_{\mathrm{[OIII]}}=$F[OIII]$_2$/F[OIII]$_1$ gives the multiplication factor used to make the narrow-band [OIII] images to include the sum of the [OIII]$\lambda$5007\AA\,+$\lambda$4959\AA\, emission line fluxes.

In summary, the  flux of each emission line ([OIII] and [NII]+H$\alpha$) was obtained as F=F$^*$-S$_C$\,F$_{C^*}$, where F$^*$ is the flux before the continuum subtraction, F is the flux after the subtraction, and F$_{C^*}$ is the continuum flux before the multiplication by the S$_C$ scale. In the case of the [OIII] image, the emission-line flux was in addition multiplied by r$_{[OIII]}$.

The resulting flux uncertainty per pixel in each line {l} is the quadrature sum of each contribuition  $\mathrm{\sigma_l^2= \sigma_{l^*}^2 + (S_{C,l}\,\sigma_{C^*})^2 + (\sigma_{C_{C,l}}\,F_{C^*})^2}$, where $\mathrm{\sigma_{l^*}}$ and $\mathrm{\sigma_l}$  are uncertainties before and after the continuum subtraction, $\mathrm{\sigma_{S_{C,l}}}$ is the continuum scale uncertainty, and $\mathrm{\sigma_{C^*}}$ is the continuum uncertainty. Both $\mathrm{\sigma_{l^*}}$ and $\mathrm{\sigma_{C^*}}$ are adopted as 3$\mathrm{\sigma_{sky}}$ of the respective image. In the case of $l$=[OIII], the uncertainty in r$_{\mathrm{[OIII]}}$ was also propagated. When a flux integration is performed, the errors of each pixel are added in quadrature.

\begin{figure}[htb!]
\includegraphics[width=1.\linewidth,angle=-0]{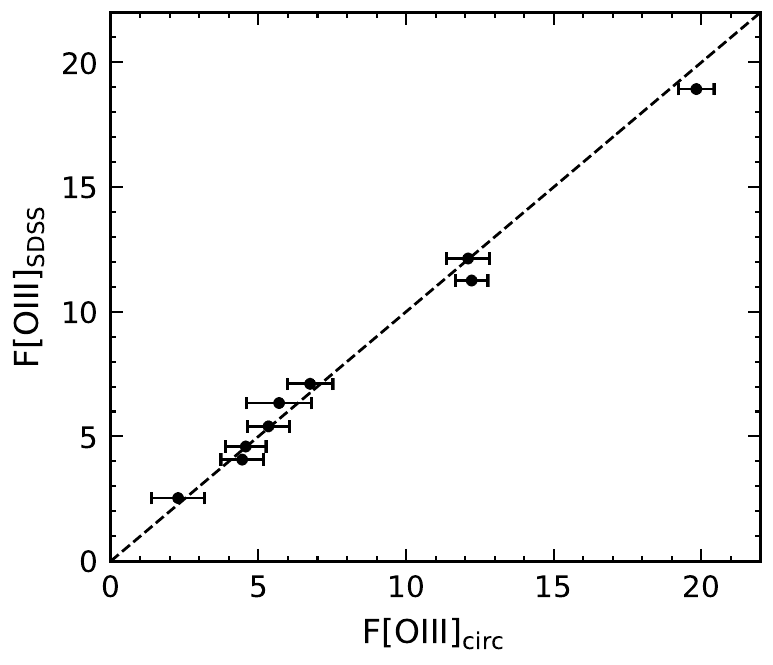}
\caption{Relation between the flux F[OIII]$_{SDSS}$ obtained from the SDSS spectra and  F[OIII]$_{circ}$,  obtained from a circular aberture of 3$^{\prime\prime}$ in our HST [OIII] images. The dashed line corresponds to unity.}
\label{fig:fluxcomp_nolog}
\end{figure}

In order to check the flux calibration after the above corrections, we integrated the resulting [OIII] HST  image flux over the SDSS circular aperture (F[OIII]$_{circ}$) and compare it with  F[OIII]$_{\mathrm{SDSS}}$, the flux of both [OIII] emission lines in the SDSS spectra. 
These values are displayed in Table\,\ref{tab:measurements}, and a comparison between them is shown in Figure \ref{fig:fluxcomp_nolog}, along with a dashed line showing the loci of equal values. It can be seen that the fluxes agree with each other within the uncertainties, supporting the robustness of our reduction and calibration processes.


The bandwidths ($\Delta_l$) of each filter were obtained from Astrolib PySynphot. It corresponds to the width of a box-like throughput curve, with the same area of the bandpass profile and heigth equal to the throughput at the average wavelength of the bandpass (\textit{avgwave} in PySynphot). The continuum and the [OIII] and H$\alpha$+[NII] narrow-band images were multiplied by the corresponding bandwidth when fluxes in units of $\mathrm{erg\,cm^{-2}\,s^{-1}}$ were required.

\begin{deluxetable}{cccccc}
\tablecolumns{6} 
\tablewidth{0pt} 
\tablecaption{Observation log\label{tab:observation}}
\tablehead{
ID & Name & Date & Filter & Exptime & Region \\
(1)  & (2)  & (3)    & (4)     & (5) &(6)   \\
}
\startdata
1& J082313 & 2015-01-01 & F775W  & 200           & Cont.       \\
  &       & 2014-12-31 & FR931N & 2542          & H$\alpha$+[NII] \\
  &      & 2015-01-01 & FR716N & 1866          & [OIII]          \\
2& J084135 & 2015-02-16 & FR647M & 200           & Cont.       \\
  &     & 2015-02-16 & FR716N & 2506          & H$\alpha$+[NII] \\
  &      & 2015-02-16 & FR551N & 2020          & [OIII]          \\
3& J085829 & 2015-02-08 & F775W  & 200           & Cont.       \\
   &     & 2015-02-08 & FR931N & 2600          & H$\alpha$+[NII] \\
   &     & 2015-02-08 & FR716N & 1927          & [OIII]          \\
4& J094521 & 2015-04-22 & FR647M & 200           & Cont.       \\
  &      & 2015-04-22 & FR716N & 2516          & H$\alpha$+[NII] \\
  &      & 2015-04-22 & FR551N & 2031          & [OIII]          \\
5& J110952 & 2015-02-16 & F775W  & 200           & Cont.       \\
   &     & 2015-02-16 & FR853N & 2600          & H$\alpha$+[NII] \\
   &     & 2015-02-16 & FR656N & 2029          & [OIII]          \\
6& J113710 & 2015-09-08 & F775W  & 200           & Cont.       \\
  &      & 2015-09-08 & FR931N & 2754          & H$\alpha$+[NII] \\
  &      & 2015-09-08 & FR716N & 2081          & [OIII]          \\
7& J123006 & 2015-07-04 & F775W  & 200           & Cont.       \\
  &      & 2015-07-04 & FR931N & 2562          & H$\alpha$+[NII] \\
   &     & 2015-07-04 & FR716N & 1889          & [OIII]          \\
8& J135251 & 2015-05-13 & FR647M & 200           & Cont.       \\
   &     & 2015-05-13 & FR782N & 1786          & H$\alpha$+[NII] \\
   &     & 2015-05-13 & FR601N & 1976          & [OIII]          \\
9& J155019 & 2015-06-17 & FR647M & 200           & Cont.       \\
  &      & 2015-06-17 & FR782N & 2516          & H$\alpha$+[NII] \\
   &     & 2015-06-17 & FR551N & 1875          & [OIII]          \\
\enddata
\tablecomments{
 (1) ID used in the paper
 (2) galaxy name; 
 (3) observation date; 
 (4) HST ACS  filter;
 (5) exposure time (in seconds);
 (6) spectral region where Cont. means continuum.}
\end{deluxetable}

\begin{deluxetable*}{cccccccc}
\tablecolumns{7} 
\tablewidth{0pt} 
\tablecaption{Parameters used in the calibration of the data \label{tab:reduction}}
\tablehead{
ID & Name & $\mathrm{\Delta_{C}}$ & $\mathrm{\Delta_{[NII]{+}H\alpha}}$ & $\mathrm{\Delta_{[OIII]}}$ & $\mathrm{S_{C,[OIII]}}$ & $\mathrm{S_{C,[NII]{+}H\alpha}}$ & $\mathrm{r_{[OIII]}}$ \\
(1)  & (2)                   & (3)                               & (4)                        & (5)                     & (6)                            & (7)
}
\startdata
1 & J082313 & 1454.59 & 200.44 & 136.96 & 0.46$\pm$0.09 & 0.3$\pm$0.1   & 1.16$\pm$0.02 \\
2 & J084135 & 482.21  & 139.54 & 98.78  & 0.98$\pm$0.05 & 1.06$\pm$0.04 & 1.20$\pm$0.02 \\
3 & J085829 & 1454.59 & 204.58 & 139.33 & 0.79$\pm$0.07 & 0.68$\pm$0.06 & 1.17$\pm$0.02 \\
4 & J094521 & 489.11  & 141.00 & 101.72 & 0.91$\pm$0.04 & 0.89$\pm$0.04 & 1.19$\pm$0.02 \\
5 & J110952 & 1454.59 & 147.79 & 126.34 & 0.91$\pm$0.05 & 0.81$\pm$0.05 & 1.15$\pm$0.02 \\
6 & J113710 & 1454.59 & 191.44 & 131.02 & 0.71$\pm$0.09 & 0.6$\pm$0.2   & 1.17$\pm$0.02 \\
7 & J123006 & 1454.59 & 195.50 & 133.06 & 0.4$\pm$0.1   & 0.5$\pm$0.2   & 1.16$\pm$0.02 \\
8 & J135251 & 538.57  & 154.73 & 118.60 & 0.95$\pm$0.06 & 0.84$\pm$0.03 & 1.16$\pm$0.02 \\
9 & J155019 & 495.44  & 135.39 & 103.95 & 1.14$\pm$0.06 & 0.88$\pm$0.05 & 1.19$\pm$0.02 \\
\enddata 
\tablecomments{
 (1) galaxy identification in the paper;
 (2) galaxy name;
 (3) continuum bandwidth (in \AA); 
 (4) [NII]+H$\mathrm{\alpha}$ filter bandwidth (in \AA); 
 (5) [OIII] filter bandwidth (in \AA);
 (6) Continuum scale factor used for [OIII];
 (7) Continuum scale factor used for [NII]+H$\mathrm{\alpha}$;
 (8) Correction factor for the partial coverage of [OIII]$\lambda$4959 line by the [OIII] filter bandpass.}
\end{deluxetable*}

\section{Results}\label{sec:results}

As we have used the SDSS spectra to correct our continuum images from the contribution of the emission-lines as well as to correct for the usually partial coverage of the [OIII]$\lambda$4959 line emission, we begin by presenting below the results of the measurements of the SDSS integrated spectra (corresponding to an aperture of 3$^{\prime\prime}$ diameter.

\subsection{SDSS spectra emission-line fluxes} \label{SDSS_fluxes}

The emission lines of the SDSS spectra were fitted using fourth-order Gauss-Hermite polynomials. In the case of the [OIII] $\lambda$5007 and $\lambda$4959 emission lines, the same radial velocity and velocity dispersion were adopted, as well as the same $h_3$ and $h_4$ coefficients, and the two lines were constrained to have a flux ratio F[OIII]$\lambda5007$/F[OIII]$\lambda4959=3$ \citep{ost06}. When one component was not enough to fit the line profile, additional components were included (keeping the same constrains above between them). The fits were obtained via Chi-Square minimization using the SLSQP Method available at Python library SciPy. The same procedure was adopted in the fit of the [NII] doublet lines. H$\beta$ was fitted using one or more Gauss-Hermite profiles. 

Objects 4, 5 and 8 show a broad base in the [NII]+H$\alpha$ emission lines, which could indicate the presence of a broad H$\alpha$ component. However, we were able to fit these profiles similarly well without the need of a broad H$\alpha$  component, by assuming that both the H$\alpha$ and [NII] profiles match that of [OIII]$\lambda$5007 in velocity space, with the broad base being a consequence of the merging of wings in the narrow emission lines. We opted for adopting this solution as the most physically compelling, even though a solution including a broad H$\alpha$ component cannot be discarded. The total fluxes and widths of the emission lines in the SDSS spectra are displayed in Table\,\ref{tab:sdssfit}. Uncertainties in the fit of the [OIII]$\lambda$5007 emission line are always below 2.5\%, while for the fainter lines, uncertainties reach at most 5\%. 

We also include, in the 9$^{th}$ column of Table\,\ref{tab:sdssfit}, the value of the ratio $\eta$=H$\alpha$/([NII]$+$H$\alpha$) obtained from the SDSS spectra, that we have used to estimate the contribution of H$\alpha$ to the narrow-band HST images that contain both the H$\alpha$ and [NII] emission lines. In the cases of the galaxies for which we could not obtain $\eta$ from the SDSS spectra (because these lines are beyond the observed spectral range), we have used the average of the values obtained for the other galaxies, and list these values in boldface in the table.

In the last column of Table\,\ref{tab:sdssfit}), we list the velocity v$_{80}$=W$_{80}/1.3$, where W$_{80}$ is the width of the [OIII]$\lambda$5007 emission-line profile that comprises 80\% of the line flux. This will be used in Section\,\ref{feedback} as an estimate of the velocity of the gas outflow in the NLR, as it probes the highest velocity gas that contributes to the emission in the profile wings.




\begin{deluxetable*}{cccccccccc}
\tablecolumns{5} 
\tablewidth{0pt} 
\tablecaption{SDSS spectra measurements \label{tab:sdssfit}}
\tablehead{
ID  & Name    & $\mathrm{F(H\beta)}$ & $\mathrm{F[OIII]}$ & $\mathrm{F(H\alpha)}$ & $\mathrm{F[NII]_{\lambda6584}}$ & $\mathrm{W_{[OIII]}}$ & $\mathrm{W_{H\alpha}}$  & $\mathrm{\eta}$ & $\mathrm{v_{80}}$ \\
(1) & (2)     & (3)                  & (4)                & (5)                   & (6)                             & (7)                                 & (8)                      & (9)                     &(10)
}
\startdata
1   & J082313 & 0.290                & 4.595              & -                     & -                               & 375                                 & -                        & \textbf{0.571}       &411          \\
2   & J084135 & 0.761                & 12.135             & 3.296                 & 1.081                           & 396                                 & 429                      & 0.696        &336     \\
3   & J085829 & 0.196                & 2.528              & -                     & -                               & 659                                 & -                        & \textbf{0.571}       &958          \\
4   & J094521 & 1.335                & 18.93              & 5.186                 & 4.947                           & 456                                 & 467                      & 0.440         &896    \\
5   & J110952 & 0.342                & 6.34               & 1.920                 & 2.207                           & 1014                                & 570                      & 0.395         &1324   \\
6   & J113710 & 0.268                & 4.071              & 0.922                 & 0.343                           & 730                                 & 785                      & 0.668         &715   \\
7   & J123006 & 0.383                & 5.404              & -                     & -                               & 880                                 & -                        & \textbf{0.571}       &995 \\
8   & J135251 & 0.613                & 7.11               & 2.810                 & 2.158                           & 821                                 & 768                      & 0.494          &807  \\
9   & J155019 & 1.193                & 11.252             & 4.784                 & 1.122                           & 355                                 & 366                      & 0.734        &333    \\                          
\enddata 
\tablecomments{
(1) galaxy identification in the paper;
(2) galaxy name;
(3)-(6) Fluxes (in units of $\mathrm{10^{-14}\,ergs\,s^{-1}}$); in the case of [OIII], it is the sum of the $\lambda\lambda4959,5007$ emission lines;
(7)-(8) FWHM of the emission lines in units of $\mathrm{km\,s^{-1}}$;
(9) ratio $\eta=H\alpha/([NII]{+}H\alpha)$, where the bold highlight indicates that the line was missing in the SDSS spectra and the value thus corresponds to the average of the ratios obtained for the other galaxies.
 (10) $v_{80}=W_{80}/1.3$, where $W_{80}$ (in km\,s$^{-1}$) is the width of [OIII]$\lambda$5007.
}
\end{deluxetable*}

\subsection{HST  images}

The corrected continuum and emission-line images are shown in Figures\,\ref{fig:panel1}--\ref{fig:panel9}. In the top left panel we show the continuum images, in the top right the continuum-subtracted [OIII] images, in the bottom right the continuum-subtracted H$\alpha$+[NII] images and in the bottom left panel an excitation map obtained as the ratio between the [OIII] and [NII]+H$\alpha$ images. The images have been multiplied by the bandwidth, and [OIII] refers to the sum of the $\lambda$4959 and $\lambda$5007 emission lines. The white bar corresponds to 1$''$, and shows the scale of the images in kpc. The black cross marks the position of the galaxy nucleus, adopted as corresponding to the peak of the continuum flux.

The flux levels shown in the [OIII] and [NII]+H$\alpha$ images of Figures\,\ref{fig:panel1}-\ref{fig:panel9} range from 1$\sigma_{sky}$ to the maximum flux value in the galaxy image (F$_{max}$). Four equally spaced contours, ranging from 3$\sigma_{sky}$ to F$_{max}$, have been overplotted on theses images. The contours were obtained after smoothing the images by a Gaussian filter with 1 pixel standard deviation. 

In the case of the excitation maps, only pixels for which both images have values higher then 3$\sigma_{sky}$ were considered. The values of the remaining pixels were set to zero. We have introduced a contour in these maps to highlight regions (inside the contour) with the highest excitation values, typical of Sy galaxies, whiile lower values are typical of LINERs. This level was chosen at [OIII]$\lambda5007$/H$\beta=5.25$, and the conversion to [OIII]/([NII]+H$\alpha$) was calculated using the corresponding fluxes of [NII]+H$\alpha$ in the SDSS spectra (further discussed in Sec.\,\ref{excitation}). Dashed contours were used for galaxies for which the [NII]/H$\alpha$ ratio could not be measured because these lines are beyond the spectral range of the SDSS spectrum, and have thus been estimated using average data from the other galaxies.

\subsubsection{Continuum}
\label{sec:continuum}

Although the continuum images have been corrected for the {\it average} contamination of the emission lines via the S$_{C,[OIII]}$ and S$_{C,([NII]+H\alpha)}$ scale factors, the images in the continuum show some features that may be due to the contamination from the emission lines, as discussed in Sec. \ref{sec:observations}. The objects showing the largest corrections ($>$\,20\% of the continuum flux) due to emission-line contaminations are the targets 1, 7, 6 and 3 (from the largest to the lowest corrections). We discuss each of these cases below, and the possible consequences for the origin of the features seen in the continuum.

Target 1 (Fig.\,\ref{fig:panel1}) shows in the continuum image a possible companion to the south-east, although its continuum flux is low, while in the [OIII] image its emission is much stronger, even after subtracting the continuum with significant emission-line contribution.This feature is thus most probably a detached cloud of gas and not a companion galaxy. In the case of target 7 (Fig.\ref{fig:panel7}), the extended feature seen in the continuum image to the south-east is most probably also due to contamination from the [OIII] emission lines, as seen by the strong corresponding feature in the [OIII] image. In the case of target 6 (Fig.\,\ref{fig:panel6}), the extended appearance in the continuum from the south-east to the north-west may have some contamination from the strong emission seen in the [OIII] line image. A compact object to the north-west is faint in the emission-line images and is most probably a companion in this case. Finally, in the case of target 3 (Fig.\,\ref{fig:panel3}), the continuum shows what seems to be a second nucleus, as its brightness approaches that of the nucleus (adopted as the brightest peak). Due to the [OIII] emission-line contamination in the continuum filter, one may think that this feature is due to this contamination. But, if this is the case, then the [OIII] image should have a strong knot of emission at this location, even if there is some contamination from the line to the continuum. But what we see instead is even a small depression in the emission-line images at this location, which implies that there is not much [OIII] emission in this knot, and the feature in the continuum is most probably a secondary nucleus, from a possible recent capture of a small galaxy.

In the other targets, where the contamination of the continuum by the emission lines is lower than 20\% it can be concluded that: target 2 is in obvious interaction with a companion; target 4 shows some assymetry in the continuum but other than that no signature of on-going interaction; target 5 is essentially round, with no signature of interaction in the continuum; target 8 seems to be an on-going merger and target 9 shows an object 2$^{\prime\prime}$ ($\approx$\,5\,kpc in the plane of the sky) to the south that seems to be a companion galaxy.

The contamination of the emission lines in the continuum images may also cause a small decrease in the flux in the [OIII] narrow-band images and also in the H$\alpha$+[NII] images in the case ot target 5, but, as the structure of the continuum image is in most cases more compact than that of the emission-line images, it does not affect the properties of the most extended gas, and in particular the measured extent of the ENLR, one of the main results of the present paper.

\subsubsection{Emission-line images}

All galaxies show [OIII] and H$\alpha$+[NII] extended emission up to several kpc in the plane of the sky. Elongated structures clearly defining ionization axes are observed for targets 1, 2, 4, 6, 7, 8 and 9, supporting the presence of a collimating structure of the nuclear radiation. The only exceptions are targets 3 and 5, what may be due to a more ``face on'' orientation of the ionization axis.

\subsection{Excitation maps}
\label{excitation}

\begin{figure*}[htb!]
\centering
\includegraphics[width=.9\linewidth,angle=-0]{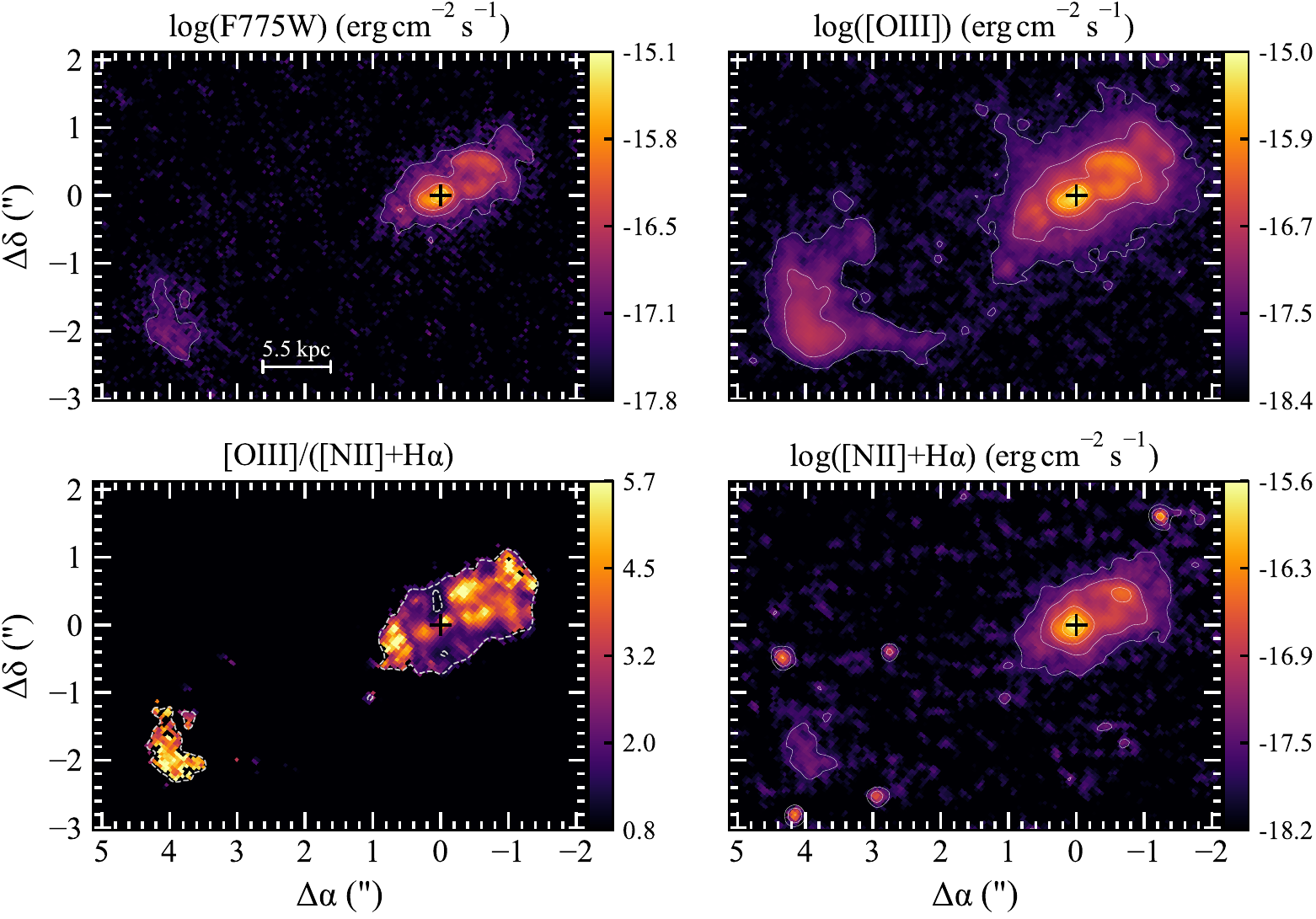}
\caption{SDSS-J082313.50+313203.7 (target 1). Top left: continuum image; top right: [OIII] narrow-band image; bottom right: H$\alpha$+[NII] narrow-band image; bottom left: excitation map. Contours in the two top panels and in the bottom right panel range from 3$\sigma_{sky}$ to the maximum value. In the bottom left panel, the contour separates the regions of highest excitation (inside the contour), corresponding to [OIII]/H$\beta=5.25$ and  [OIII]/([NII]+H$\alpha$)=1.3. North is up and East to the left in all panels. In this galaxy, the continuum image has contamination from the [OIII] emission lines, and the detached structure to the south-east may be a gas cloud appearing in the continuum due to this contamination (see text).}
\label{fig:panel1}
\end{figure*}

\begin{figure*}[htb!]
\centering
\includegraphics[width=.7\linewidth,angle=-0]{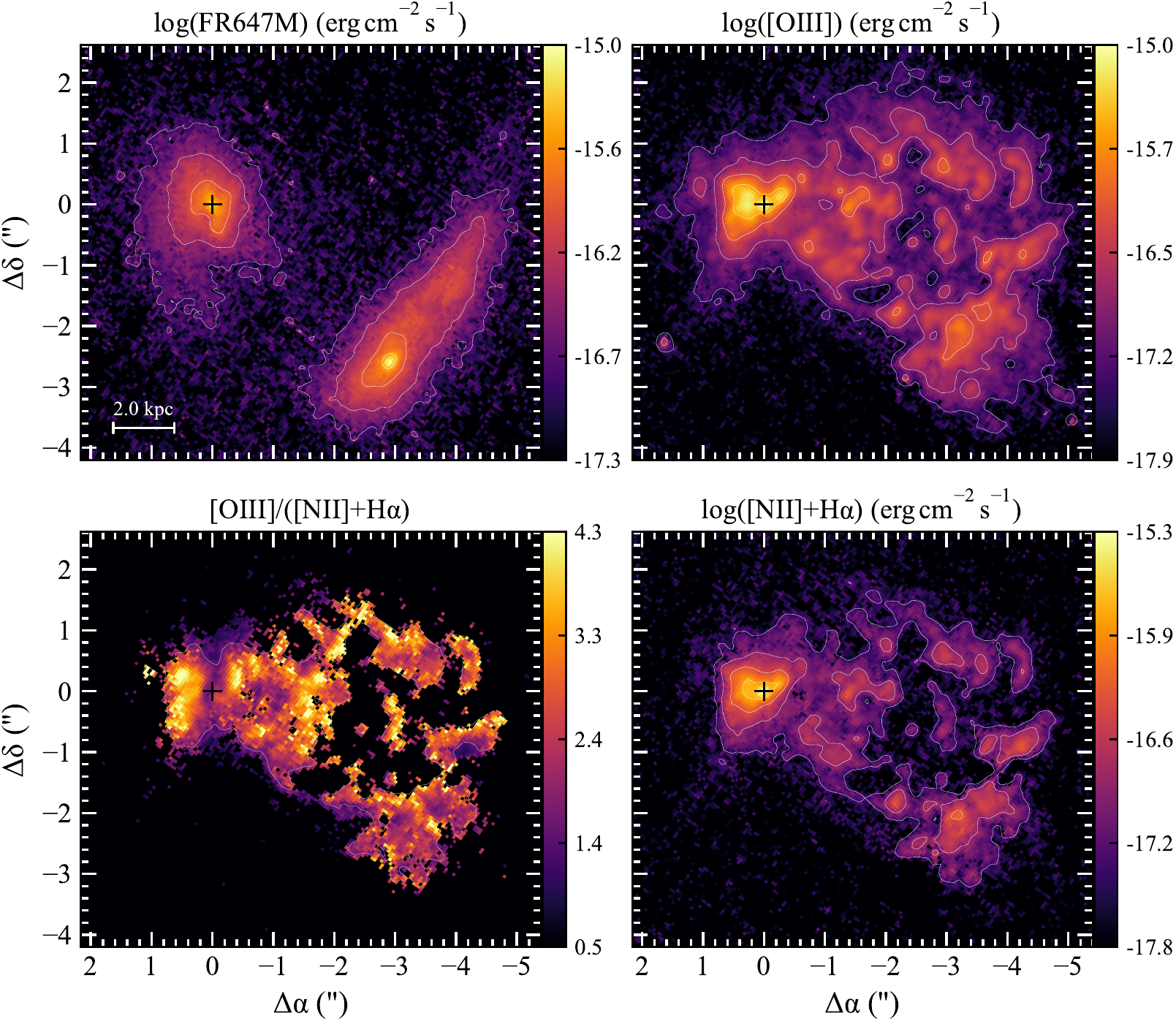}
\caption{As in Fig.\,\ref{fig:panel1} for SDSS-J084135.04+010156.3 (target 2).}
\label{fig:panel2}
\end{figure*}

\begin{figure*}[htb!]
\centering
\includegraphics[width=.6\linewidth,angle=-0]{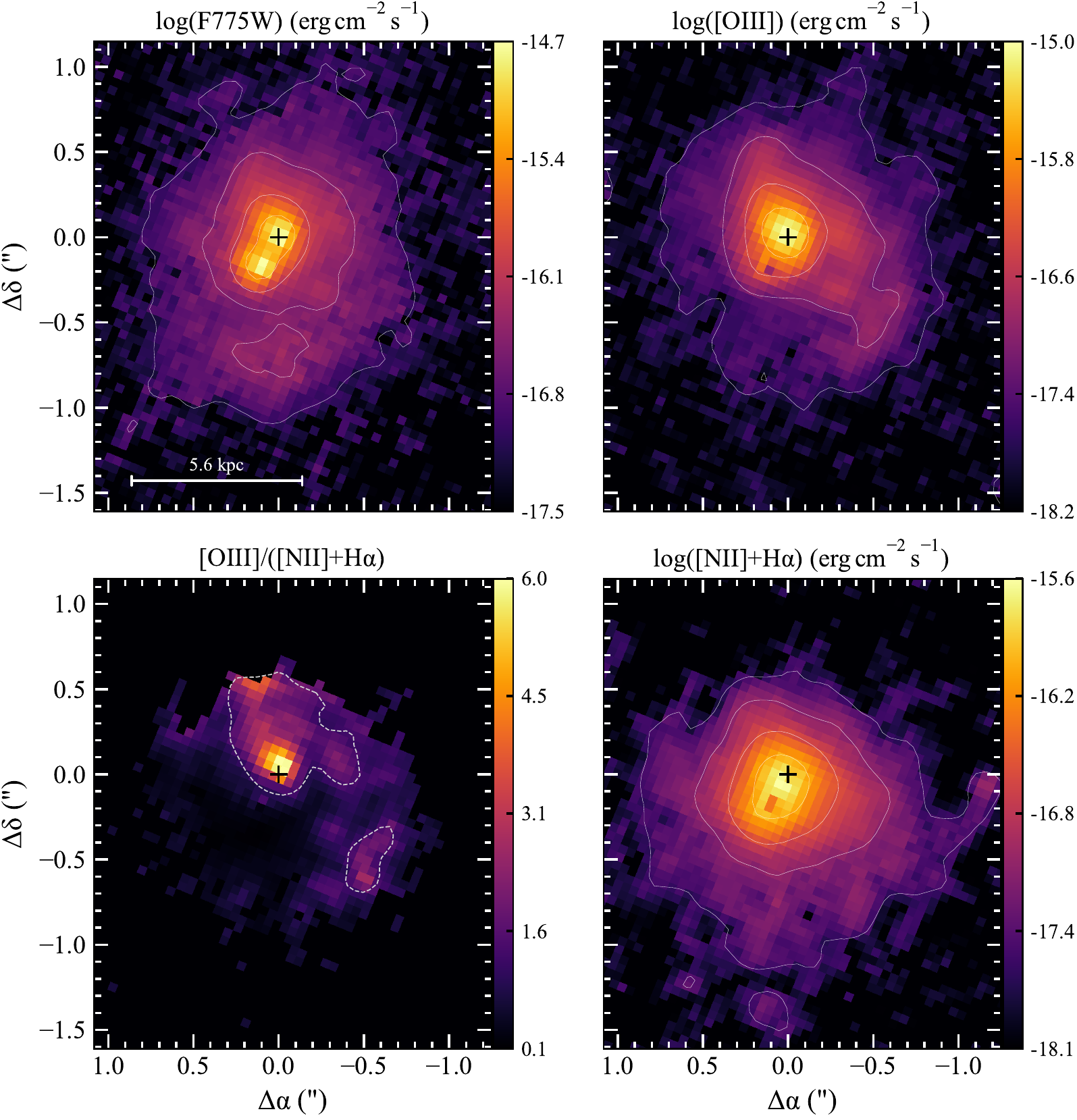}
\caption{As in Fig.\,\ref{fig:panel1} for SDSS-J085829.58+441734.7 (target 3). In this galaxy, the continuum image has contamination from the [OIII] emission lines, but the secondary nucleus seen in the continuum image is not due to this contamination (see text).}
\label{fig:panel3}
\end{figure*}

\begin{figure*}[htb!]
\centering
\includegraphics[width=1.\linewidth,angle=-0]{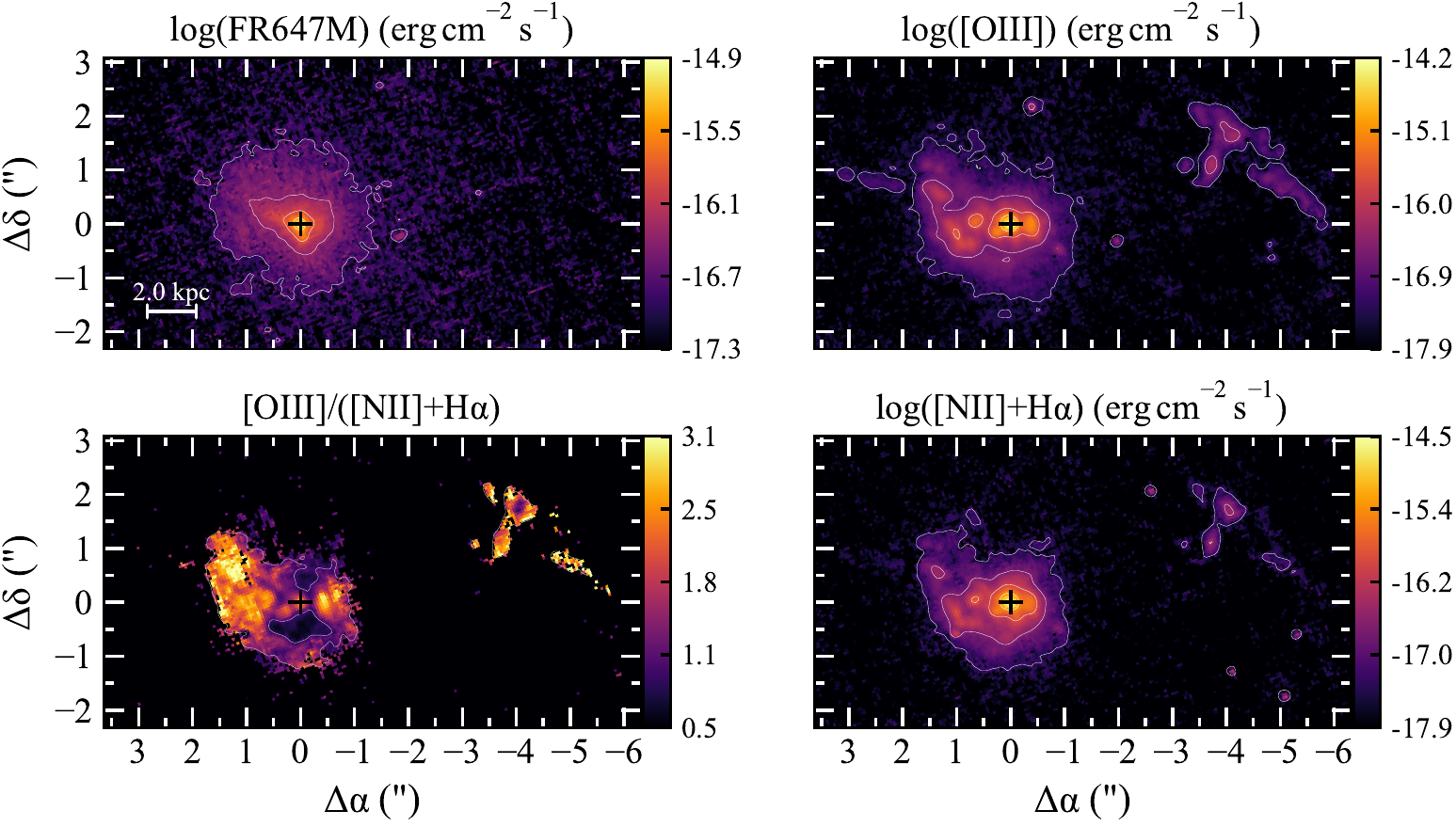}
\caption{As in Fig.\,\ref{fig:panel1} for SDSS-J094521.34+173753.3 (target 4).}
\label{fig:panel4}
\end{figure*}

\begin{figure*}[htb!]
\centering
\includegraphics[width=.5\linewidth,angle=-0]{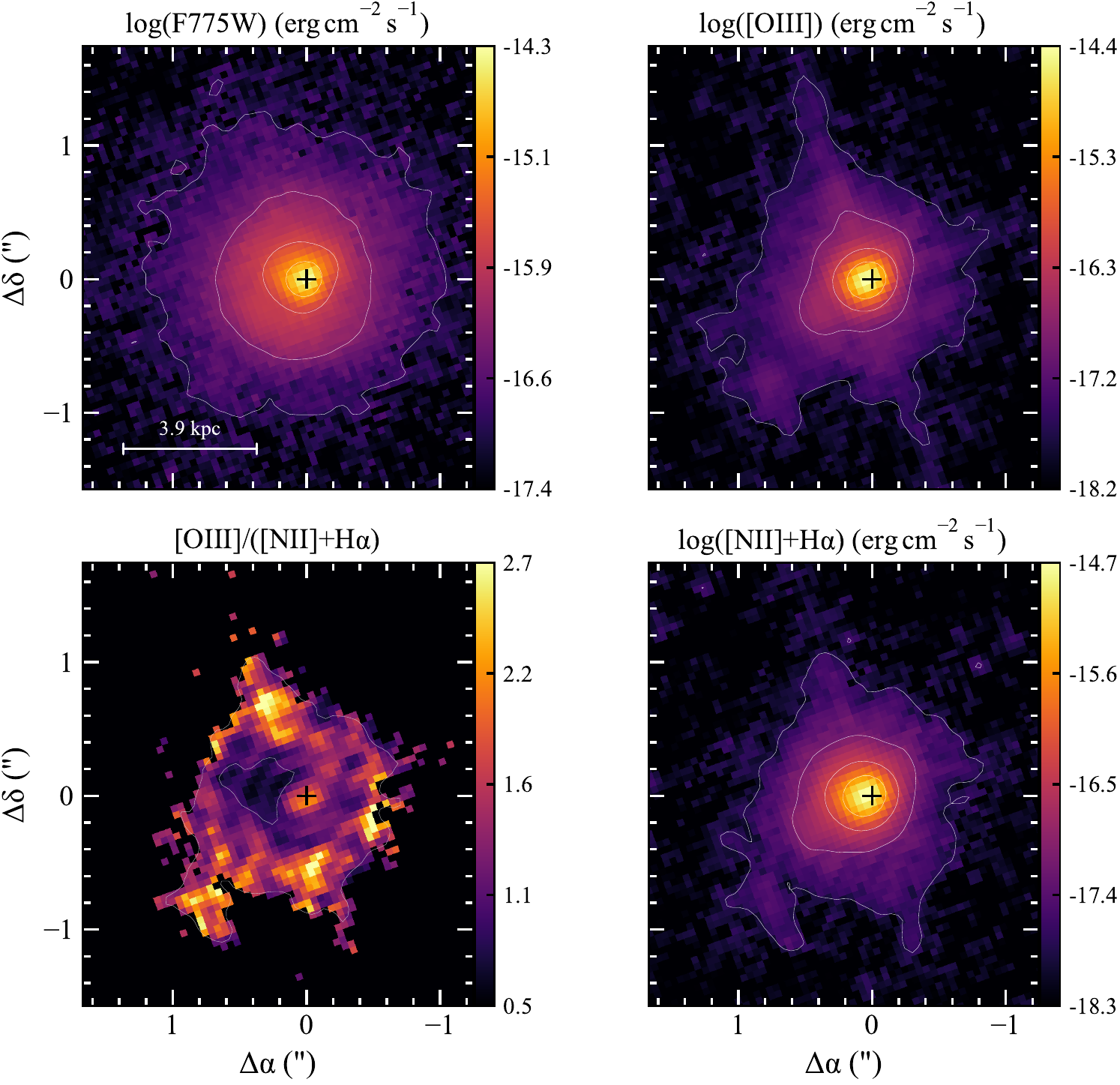}
\caption{As in Fig.\,\ref{fig:panel1} for SDSS-J110952.82+423315.6 (target 5).}
\label{fig:panel5}
\end{figure*}

\begin{figure*}[htb!]
\centering
\includegraphics[width=.7\linewidth,angle=-0]{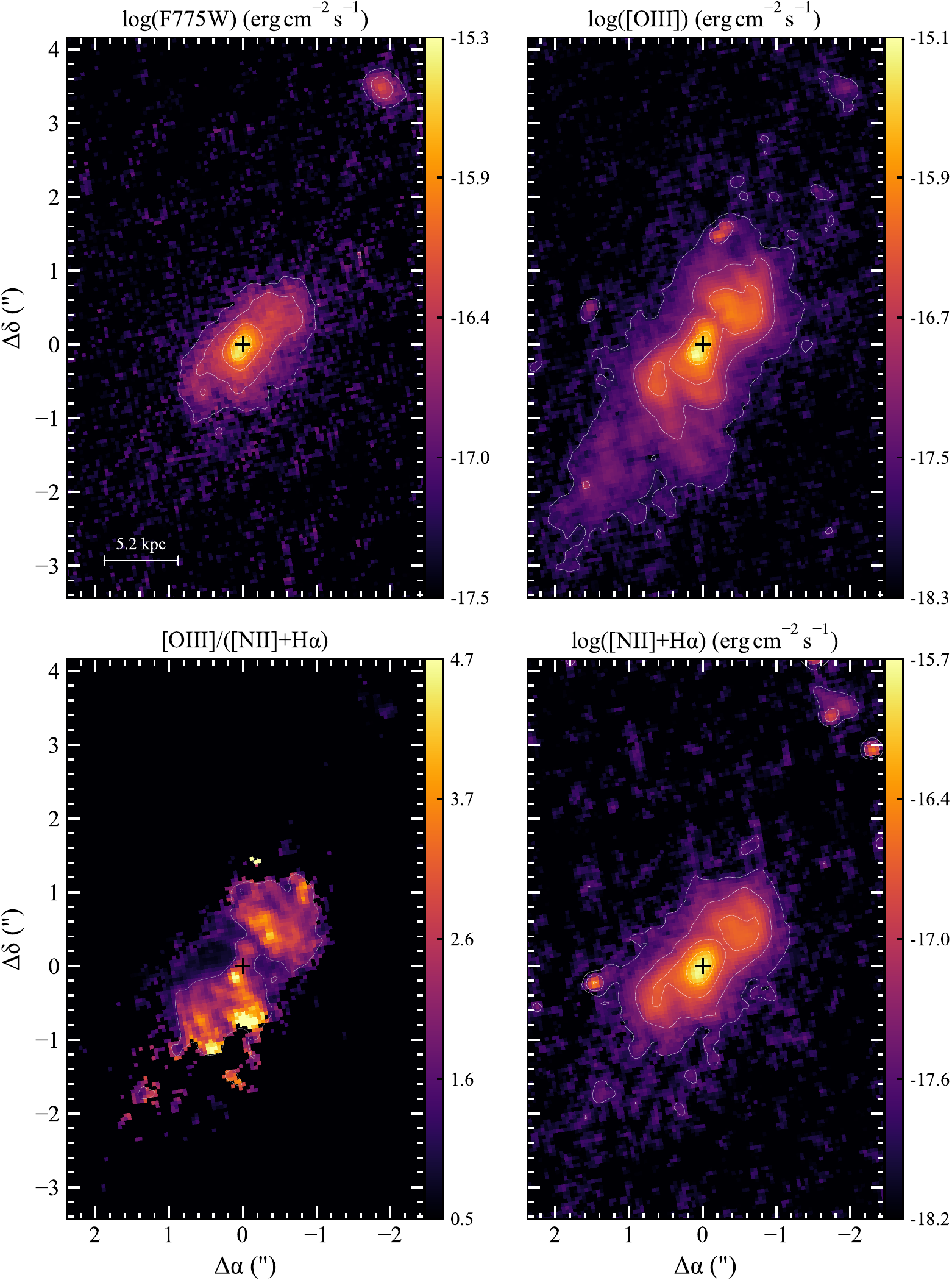}
\caption{As in Fig.\,\ref{fig:panel1} for SDSS-J113710.77+573158.7 (target 6).  In this galaxy, the continuum image has contamination from the [OIII] emission lines, but the detached structure to the north-west may be a companion as it does not appear in the emission-line images (see text).}
\label{fig:panel6}
\end{figure*}

\begin{figure*}[htb!]
\centering
\includegraphics[width=.6\linewidth,angle=-0]{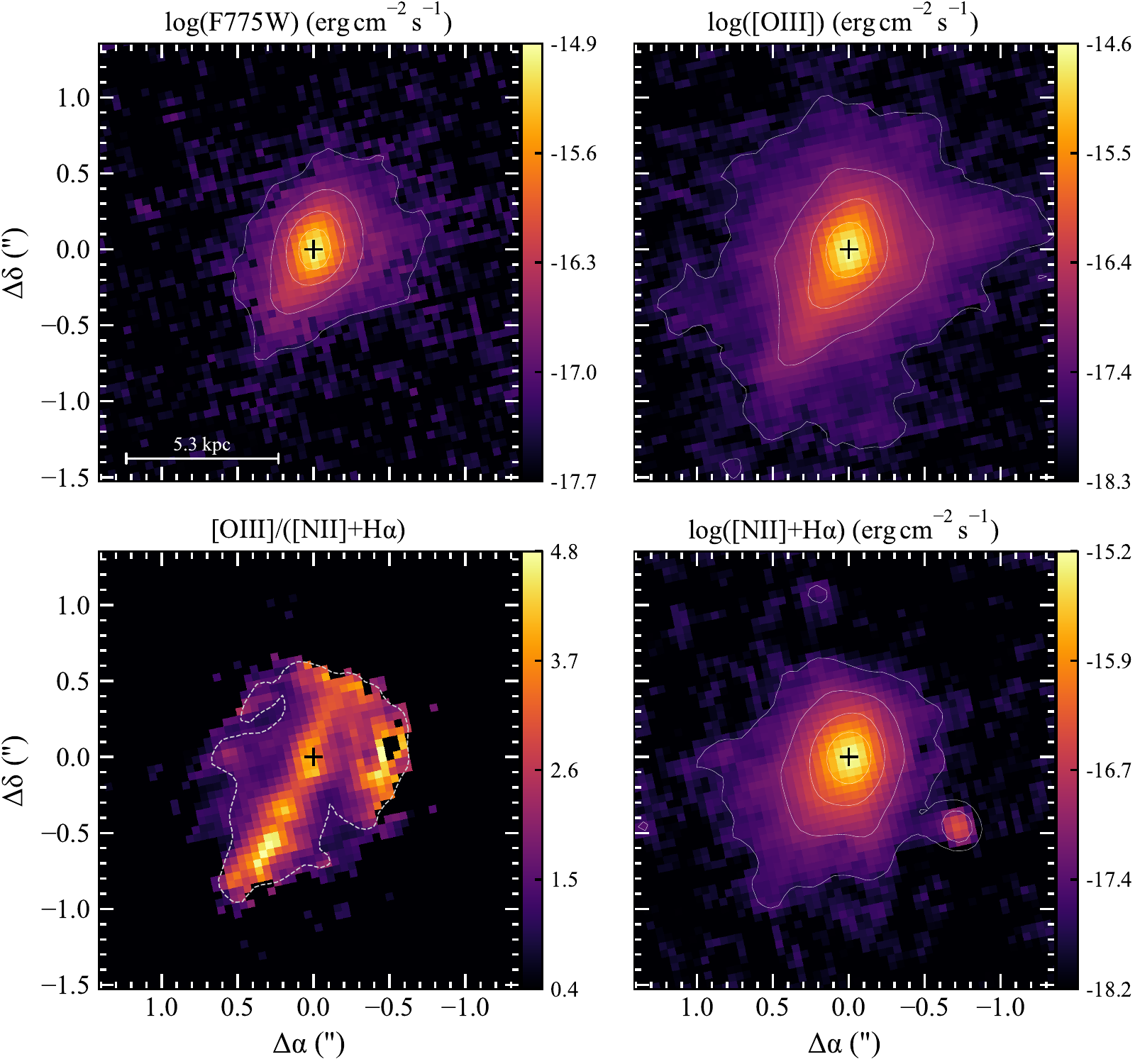}
\caption{As in Fig.\,\ref{fig:panel1} for SDSS-J123006.79+394319.3 (target 7). In this galaxy, the continuum image has contamination from the [OIII] emission lines, and the elongation to the south-east may be due to this contamination (see text).}
\label{fig:panel7}
\end{figure*}

\begin{figure*}[htb!]
\centering
\includegraphics[width=.8\linewidth,angle=-0]{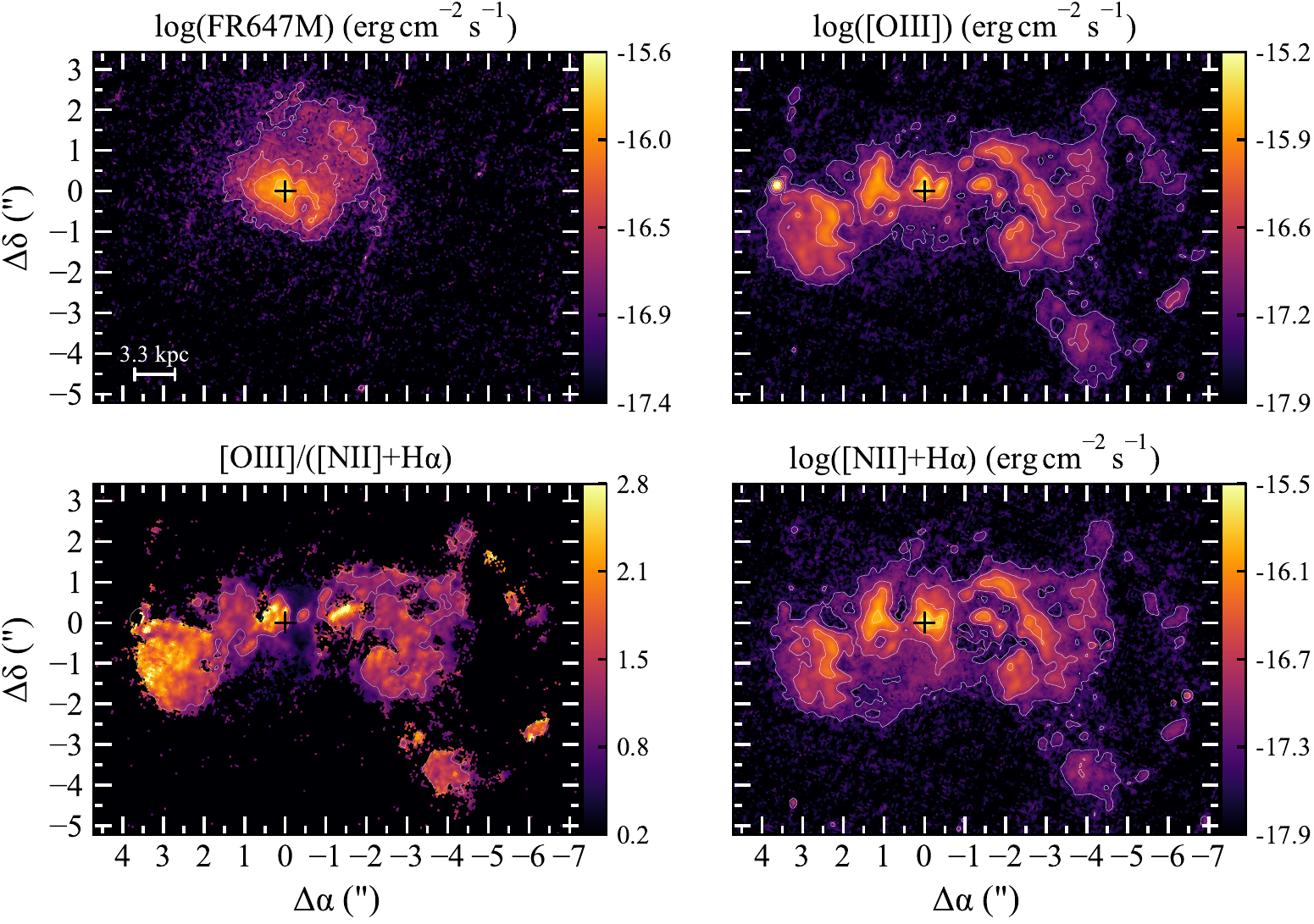}
\caption{As in Fig.\,\ref{fig:panel1} for SDSS-J135251.21+654113.2 (target 8).}
\label{fig:panel8}
\end{figure*}

\begin{figure*}[htb!]
\centering
\includegraphics[width=.8\linewidth,angle=-0]{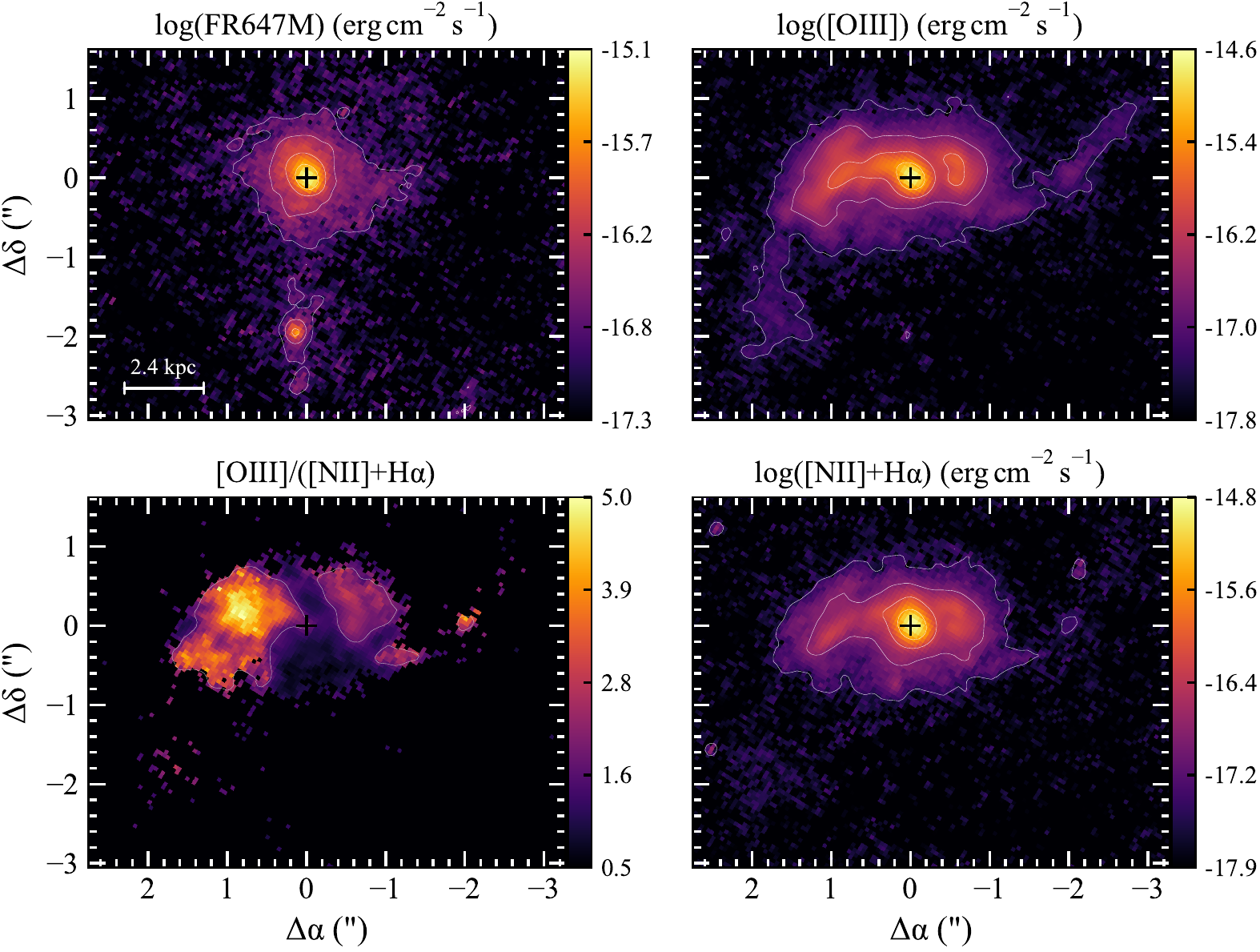}
\caption{As in Fig.\,\ref{fig:panel1} for SDSS-J155019.95+243238.7 (target 9).}
\label{fig:panel9}
\end{figure*}

The bottom-left panels of Figs.\,\ref{fig:panel1}-\ref{fig:panel9} show the excitation maps, obtained as the ratio between the [OIII] and [NII]+H$\alpha$ images. As discussed above, we added a contour to these maps as a reference for the location of the regions with the highest excitation (inside the contours). The contour adopted to separate these values corresponds to [OIII]$\lambda$5007/H$\beta=5.25$, a value close to the separation between LINERs (below this value) and Seyferts (above this value) in the  BPT \citet{vo87,bal81}  [OIII]/H$\beta$ vs. [NII]/H$\alpha$ diagnostic diagram, considering values of [NII]/H$\alpha\approx\,0.3--1$ for our sample. The conversion of this [OIII]/H$\beta$ value to [OIII]/([NII]$+$H$\alpha$)  -- the actual ratio shown in our excitation maps, was obtained under the assumptions [OIII]$\lambda5007$/[OIII]$\lambda4959=3$ and H$\alpha$/H$\beta=3$. This H$\alpha$/H$\beta$ ratio was adopted considering that, although its value is on average $\approx$\,4 in the SDSS spectra, it should be dominated by the emission closest to the nucleus where there may be some reddening, while for most of the nebulae (farther from the nucleus) it should be closer to $\approx$\,3. In any case, this contour is just a reference for the excitation level. In order to obtain the value of the [OIII]/([NII]+H$\alpha$) ratio, we have used the $\eta$ value listed in Table\,\ref{tab:sdssfit}). Considering the above, we have:
\begin{equation}
 \frac{[OIII]}{[NII]{+}H\alpha} = \frac{4\eta}{9}  \left( \frac{[OIII]_{\lambda5007}}{H\beta} \right)
 \label{eq:eta}
\end{equation}

These reference contours are shown as continuous lines for galaxies with available $\eta$ values, and as dashed lines for the galaxies with estimated $\eta$ values (shown in boldface in Table\,\ref{tab:sdssfit}). It is also important to point out that we are using a constant $\eta$ ratio througout the ENLR, determined from the SDSS spectra, although this value most probably varies from pixel to pixel. Figs.\,\ref{fig:panel1}-\ref{fig:panel9} show that the contours in most cases delineate a (patchy) approximately biconical structure for the highest ionization regions (orange) surrounded by lower ionization regions (purple). In the case of targets 1 and 7, these contours seem to be beyond the region separating the highest from the lower excitation, but the biconical structure is delineated by the orange regions surrounded by purple regions with the separation between them appearing at higher line ratios than that of the adopted contour.

\subsubsection{[OIII]$\lambda5007$ and [OIII]/H$\beta$ spatial profiles}\label{sec:o3_hb}

\begin{figure*}[htb!]
\centering
\includegraphics[width=1\linewidth,angle=-0]{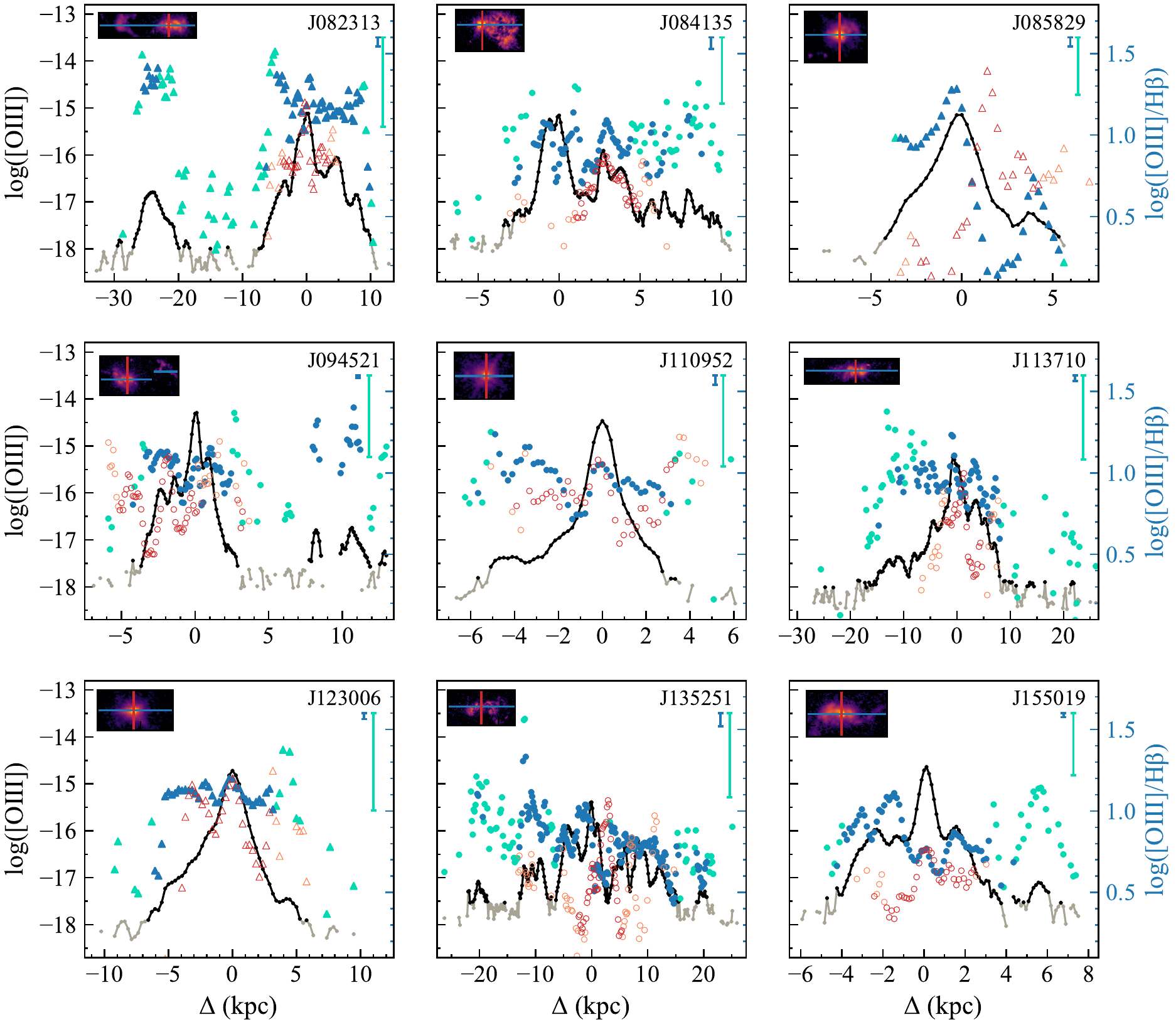}
\caption{Spatial profiles of: (1) [OIII] flux (black and grey lines); (2) [OIII]/H$\beta$ ratio (blue and torquoise symbols) along the ionization axis (blue line in the inserts at the left corner of the panels); (3) [OIII]/H$\beta$ ratio (red and orange symbols) along the direction perpendicular to the ionization axis (red line in the inserts). Black and gray lines correspond to [OIII] flux values above 3$\sigma_{sky}$ and between 3$\sigma_{sky}$ and 1$\sigma_{sky}$, respectively. Similarly, for line fluxes above 3$\sigma_{sky}$,  [OIII]/H$\beta$  line ratios are shown in blue and red, and between 3$\sigma_{sky}$ and 1$\sigma_{sky}$, in turquoise and orange. Typical error bars are shown close to the right vertical axes. Circles represent line ratios for which $\eta$ could be obtained, triangles represent cases in which an average value was used. Units are logarithmic, shown in the left axis for the [OIII] fluxes and in the right axis for the line ratios.}
\label{fig:o3hb}
\end{figure*}

In order to better quantify the excitation along the ENLR, we have used the $\eta$ values listed in Table\,\ref{tab:sdssfit} to obtain the line ratio values [OIII]$\lambda$5007/H$\beta$ (hereafter [OIII]/H$\beta$), and construct one-dimensional spatial profiles for the [OIII] fluxes and [OIII]/H$\beta$ ratios through the nucleus along the ionization axis and perpendicular to it within pseudo slits with a width corresponding to 0\farcs05 ($\approx$\,100$-$300\,pc at the galaxies).These one-dimensional profiles are shown in Figure\,\ref{fig:o3hb}. The [OIII] profiles are only shown along the ionization axis.


\subsection{Extent of the ENLR}

\begin{figure*}[htb!]
\includegraphics[width=1.\linewidth,angle=-0]{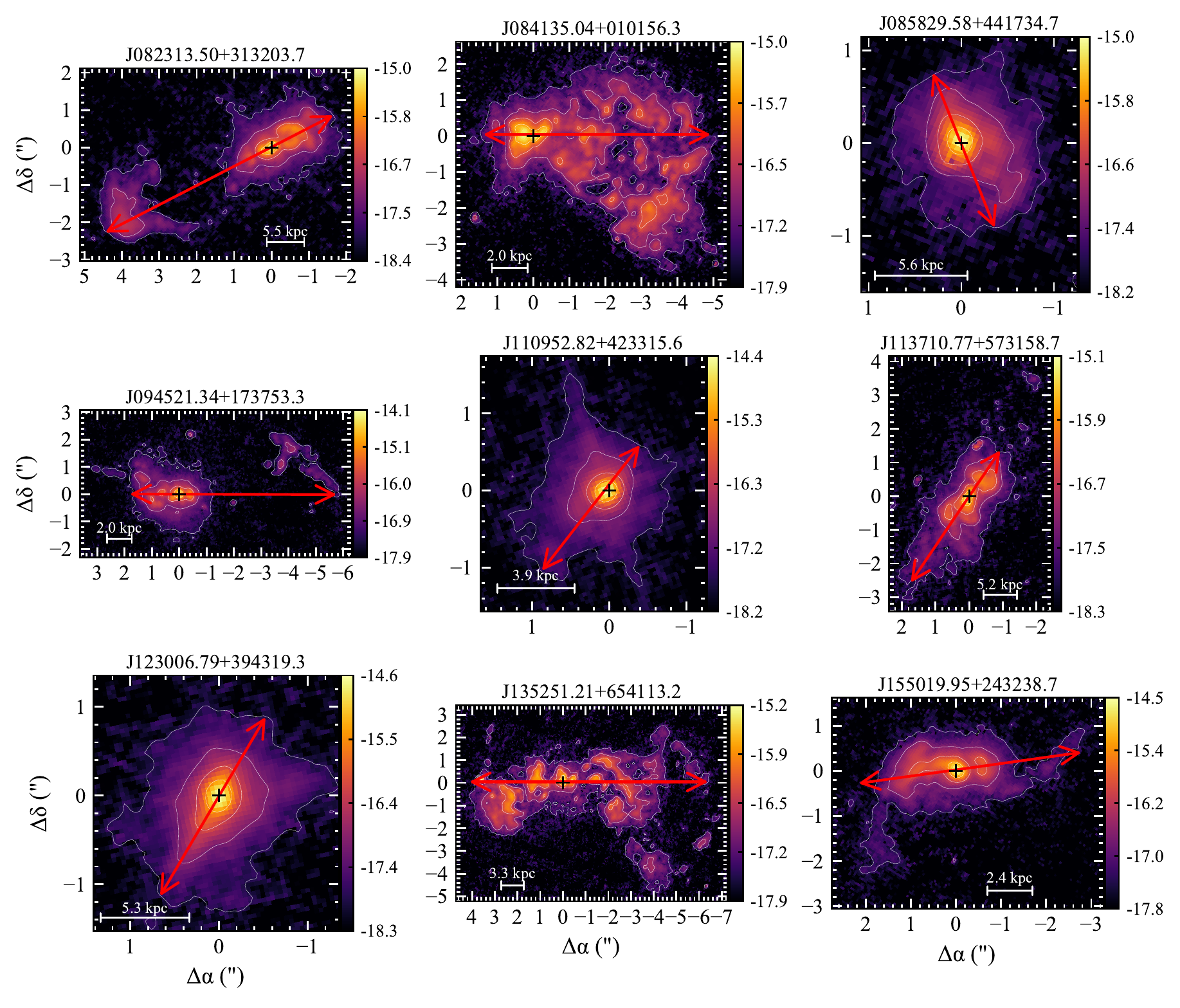}
\caption{The adopted extent of the ENLR is shown with a red arrow in the [OIII] images of the galaxies. R$_{maj}$ was adopted as the linear size corresponding at the galaxy to half the extent of the red arrow. The orientation chosen to measure the extent corresponds to the approximate ionization axis of the bicone as observed in the excitation maps of the Figures\,\ref{fig:panel1}-\ref{fig:panel9}  and listed as P.A. in Table\,\ref{tab:measurements}. The contour values range from 3$\sigma_{sky}$ to $F_{max}$.}
\label{fig:size1}
\end{figure*}

We have measured the total extent of the ionized gas emission in the [OIII] images along the ionization axis, up to an emission level corresponding to the 3$\sigma_{sky}$ contour. The  $\sigma_{sky}$ value of the [OIII] image is listed in the last column of Table\,\ref{tab:measurements}. The orientation of the ionization axis -- also listed under PA in Table\,\ref{tab:measurements} -- was adopted as that corresponding to the longest axis of symmetry of the excitation maps in Figures\,\ref{fig:panel1}-\ref{fig:panel9}. Figure\,\ref{fig:size1} shows the adopted extents over the [OIII] images for each galaxy. The uncertainty in this extent was estimated as the difference between the values measured using the contours at 2$\sigma_{sky}$ and 4$\sigma_{sky}$.

The radius of the ENLR -- which we hereafter call $R_{maj}$ -- was adopted to be half the value in kpc corresponding to  the above angular extents \citep[following][]{sch03a}, with an uncertainty propagated from those in the angular extent and corresponding galaxy scale. The R$_{maj}$ values are listed in Table\,\ref{tab:measurements} and range from 4 to 19 kpc.

\subsection{Emission-line luminosities and gas masses}

The total emission-line luminosities L[OIII] and L([NII]$+$H$\alpha$) as derived from the images were obtained by integrating the flux values above $3\sigma_{sky}$ using the distances listed in Table\,\ref{tab:sample}. The resulting  values are listed in Table\,\ref{tab:measurements}, where L[OIII]$_{\lambda5007}$ was obtained from the images after multiplying them by 0.75 in order to eliminate the contribution of the [OIII]$\lambda$4959 emission line.


\begin{deluxetable*}{cccccccccc}
\tablecolumns{10} 
\tablewidth{0pt} 
\tablecaption{Measurements from the HST images \label{tab:measurements}}
\tablehead{
\#  & Name    & $\mathrm{F[OIII]_{circ}}$ & $\mathrm{L[OIII]_{\lambda5007}}$ & PA  & $\mathrm{\theta}$ & $\mathrm{R_{maj}}$ & $\mathrm{L(H\alpha{+}[NII])}$ & $\mathrm{M_{ENLR}}$     & $\sigma_{sky,[OIII]}$       \\
(1) & (2)     & (3)                       & (4)                              & (5) & (6)               & (7)                 & (8)                           & (9)   &(10)
}
\startdata
1   & J082313 & 4.57$\pm$0.69             & 25.0$\pm$4.3                     & 117 & 6.87$\pm$0.13     & 18.76$\pm$0.93      & 8.9$\pm$8.4                   & 1.2$\pm$1.2   &0.35  \\
2   & J084135 & 12.10$\pm$0.73            & 5.37$\pm$0.65                    & 90  & 6.33$\pm$0.34     & 6.18$\pm$0.45       & 2.38$\pm$0.31                 & 0.398$\pm$0.052 &1.1 \\
3   & J085829 & 2.28$\pm$0.90             & 11.6$\pm$4.4                     & 22  & 1.80$\pm$0.14     & 5.05$\pm$0.46       & 10.8$\pm$4.3                  & 1.48$\pm$0.59 & 0.50  \\
4   & J094521 & 19.83$\pm$0.60            & 6.47$\pm$0.61                    & 90  & 7.58$\pm$0.19     & 8.41$\pm$0.47       & 4.70$\pm$0.45                 & 0.496$\pm$0.048 & 1.1 \\
5   & J110952 & 5.7$\pm$1.1               & 8.2$\pm$1.6                      & 142 & 2.07$\pm$0.14     & 4.05$\pm$0.34       & 7.4$\pm$2.0                   & 0.70$\pm$0.19 & 0.56  \\
6   & J113710 & 4.45$\pm$0.72             & 18.4$\pm$3.3                     & 146 & 4.75$\pm$0.45     & 12.3$\pm$1.3        & 9.8$\pm$5.8                   & 1.57$\pm$0.93  &0.45 \\
7   & J123006 & 5.34$\pm$0.71             & 21.8$\pm$3.4                     & 149 & 2.34$\pm$0.22     & 6.15$\pm$0.64       & 9.9$\pm$8.9                   & 1.4$\pm$1.3   & 0.42  \\
8   & J135251 & 6.75$\pm$0.77             & 18.8$\pm$2.0                     & 90  & 10.52$\pm$0.46    & 17.1$\pm$1.2        & 17.1$\pm$1.8                  & 2.03$\pm$0.22  & 1.1 \\
9   & J155019 & 12.22$\pm$0.55            & 4.83$\pm$0.51                    & 98  & 4.99$\pm$0.69     & 6.04$\pm$0.89       & 3.37$\pm$0.37                 & 0.594$\pm$0.066 & 1.2 \\
\enddata
\tablecomments{
 (1) identification number in the paper;
 (2) galaxy name
 (3) [OIII] flux measured in the HST image within an aperture of 3$''$ (in units $\mathrm{10^{-14}\,erg\,s^{-1}}$);
 (4) total [OIII]$\lambda5007$ luminosity, integrated above 3$\sigma_{sky}$ (in units $\mathrm{10^{42}\,erg\,s^{-1}}$);
 (5) PA of ionization axis (in $\mathrm{^{\circ}}$);
 (6) ENLR angular extent (in $\mathrm{''}$); 
 (7) ENLR radius (in kpc);
 (8) H$\alpha$+[NII] total luminosity, integrated above 3$\sigma_{sky}$ (in units $\mathrm{10^{42}\,erg\,s^{-1}}$);
 (9) total ionized gas mass (in units $\mathrm{10^8\,M_{\sun}})$;
(10) standard deviation of the sky value in the [OIII] image in units of $\mathrm{10^{-18}\,erg\,cm^{-2}\,s^{-1}}$.}
 \end{deluxetable*}

\subsubsection{Total ionized gas masses}

We have estimated the total ionized gas mass of the ENLR from L([NII]$+$H$\alpha$), for case B recombination \citep{ost06} following \citet{pet97}. The total luminosity of H$\beta$, originating from clouds within a total volume V$_{c}$, is $L(H\beta)=n_{e}n_{p} \, \alpha_{H\beta}^{\rm eff} \, h\nu_{H\beta} V_{c}$, where $\alpha_{H\beta}^{\rm eff}$ and $\nu_{H\beta}$ are the effective recombination coefficient and the rest frequency of H$\beta$, respectively, and $n_e$ and  $n_p$ are the number densities of electrons and protons. We consider completely ionized hydrogen clouds, thus $n_e\,{\approx}\,n_p$. The H$\alpha$ luminosity can be written as $L(H\alpha)\,{=}\,(j_{H\alpha}/j_{h\beta})\,L(H\beta)$, where $j_{H\alpha}/j_{h\beta}$ is the ratio between H$\alpha$ and H$\beta$ emissivities. Assuming L(H$\alpha$)=$\eta$\,L([NII$+$H$\alpha$]) and the same density for all clouds -- $n_{p}m_{p}$, where $m_p$ is the proton mass, the total ionized mass is $M_{ENLR}=(n_{p}m_{p})V_{c}$. Using equations above:

\begin{eqnarray}
M_{ENLR} &=& \frac{m_{p} \, \eta \, L([NII]{+}H\alpha)} {n_{e} \, (j_{H\alpha}/j_{h\beta}) \, \alpha_{H\beta}^{eff}\,h\nu_{H\beta}} \\ 
         &=& 0.238 \, \eta \, L_{42}([NII]{+}H\alpha) \, \,M_{\sun},
\end{eqnarray}\label{eq:mass}

\noindent where $L_{42}([NII]{+}H\alpha)$ is in units of $10^{42}erg\,s^{-1}$. As we do not have resolved density values from our images, we have adopted a constant density of 100\,cm$^{-3}$ thoughout the ENLR as a compromise. The value of the density should be higher in the inner kpc or so (thus we are overestimating the mass for this region) but should be lower for the outer kpcs (thus we are underestimating the mass for the external regions).
The values for the emission coefficient and recombination ratio have been obtained from \citet{ost06} for $n_e=100$\,cm$^{-3}$ and T$=10^4K$. The resulting M$_{ENLR}$ masses are shown in the 9$^{th}$ column of Table\,\ref{tab:measurements}.


\subsubsection{Ionized gas mass profiles}

\begin{figure*}[htb!]
\includegraphics[width=1.\linewidth,angle=-0]{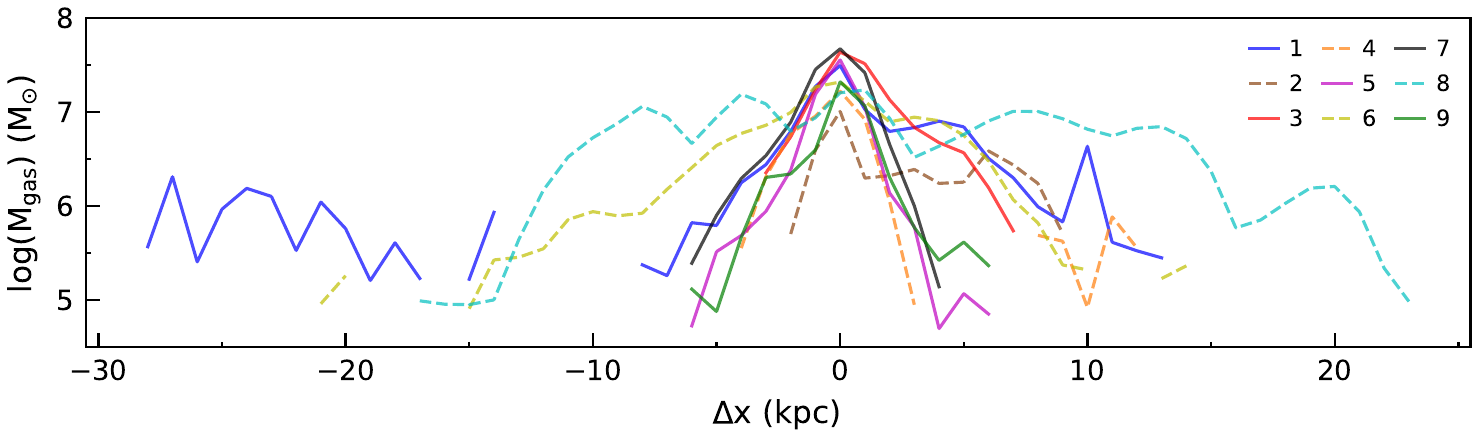}
\caption{Cummulative gas mass profiles obtained by binning the masses into 1 kpc$^2$ at the galaxies, rotating the surface mass density distributions so that the ionization axis is oriented along the x-axis and summing the mass contributions along the perpendicular direction within each kpc.}
\label{fig:surfacemass}
\end{figure*}

We can use the ionized gas masses per pixel calculated as above to obtain the ionized gas surface mass densities. As we have adopted a constant gas density, the corresponding maps are very similar to those of the H$\alpha$+[NII]. We show instead a one-dimensional profile for the mass distribution binning the data within square regions corresponding to 1\,kpc$^2$ at the galaxies. We then built a cummulative mass profile along the ionization axis by adding all masses perpendicularly to this axis within each kpc. These one-dimensional cummulative mass distributions, shown in Fig.\,\ref{fig:surfacemass}, present a ``central peak" that reaches values of 10$^{7.5}$\,M$_\odot$ in the inner kpc and decrease down to 10$^{5}$\,M$_\odot$ at about 5\,kpc from the nucleus for targets 4, 5, 7 and 9, that show an approximately symmetric mass distribution relative to the nucleus. Assymmetric (more extended to one side) and more extended mass distributions are observed for the remaining objects, whose values of cummulative masses within each kpc along the ionization axis range from $\sim$ 10$^{6}$\,M$_\odot$ to $\sim$ 10$^{7}$\,M$_\odot$ at distances between 10 and 20\,kpc. Targets 1, 6 and 8 show the most extended mass distributions. All objects with assymetric mass profiles correspond to galaxies with signatures of interactions, the only exception being target 9 that, although seeming to have a symmetric mass profile, shows a companion.



\section{Discussion}\label{discussion}


One recent development in the study of AGN was the realization that AGN flicker on scales $\le\,10^5$ yr \citep[e.g.][]{novak11,hickox14}, what is supported by observations of even our home galaxy by the discovery of the ``Fermi bubbles" in the vicinity of the galactic center \citep{su10}. Further evidence of this past activity of the galaxy center has also been seen in the form of a ``fossil imprint" of a past powerful flare (1--3\,Myr ago) on the Magellanic Stream at a distance of 50--100\,kpc from the galactic center \citep{bh13}. In addition, two absorbing structures with velocities of $-235$ and $+250$\,km\,s$^{-1}$ in the light of background quasars have been recently observed between the galactic nucleus and these ionized regions of the Magellanic Stream \citep{fox15}, in line with an origin in the front and back walls of a bipolar outflow due this past nuclear activity.

In light of this known intermitency of the nuclear activity in galaxies, it is somewhat unexpected that the line emission from the ENLR is as continuous as observed in our sample. Eventual patchiness can be attributed to irregularites in the mass distribution, but there are no clear signatures of light echoes. An explanation for this has been previously proposed \citep{crenshaw03,sb10}: due to the low gas density of the most external regions of the ENLR ($\le$\,1\,cm$^{-3}$), the recombination time becomes of the order of the flicker time of $\approx$\,10$^{5-6}$\,yr. In any case, a time-dependent analysis of the AGN photoionisation is something we plan to do in the near future \citep[e.g.][]{bh13} using resolved spectroscopy, that is being acquired for the galaxies of our sample.

\subsection{The Ionization Cones}

Inspection of Figs.\,\ref{fig:panel1}-\ref{fig:panel9} shows that most narrow-band images present an elongated morphology indicating collimation of the ionizing radiation. The exceptions are targets 3 and 5, that show a more compact gas distribution. One possibility is that the orientation of the ionization axis is closer to ``face-on" in these cases. In the case of target 5 there is evidence that this is indeed the case. This target shows the broader base in the H$\alpha+$[NII] of the SDSS spectrum. Although we favored a fit without a broad H$\alpha$ component, the data is also consistent with the presence of such component within the fitting uncertainties, in which case the observation of a Seyfert 1 nucleus would be consistent with a more pole-one visualization of the AGN, explaining the ``rounder" ENLR in this galaxy. A ``pole-on" view of the ENLR and the possible associated outflow is also consistent with the high value of v$_{80}\sim\,1300$\,km\,s$^{-1}$, listed in Table\,\ref{tab:sdssfit} for this galaxy, possibly due to a high outflow velocity component, as expected if the direction of the outflow is close to the line-of-sight. 

The excitation maps in Figs.\,\ref{fig:panel1}-\ref{fig:panel9} reveal ionization bicones in 6 of our 9  targets, namely targets 1, 2, 4, 6, 8 and 9,  with the highest excitation levels -- with [OIII]/(H$\alpha$+[NII]) in the range $\approx$\,4--6 -- shown as orange in the figures -- closer to their central axes, surrounded by lower excitation levels -- with [OIII]/(H$\alpha$+[NII])\,$\le$\,1 -- shown as violet in the figures. Lower excitation is also observed surrounding the apex of the cones, perpendicularly to the ionization axis, supporting an obscuration of the nucleus along our line of sight due to a collimating structure -- presumably the dusty torus postulated in the Unified Model \citep{ant93}. 

Our results thus do not support the disappearence of the torus for high-luminosity AGN as postulated by some studies \citep[e.g.][]{elitzur14} at least up to luminosities of L[OIII]$=2.5\times10^{43}$\,erg\,s$^{-1}$ (the highes AGN luminosity in our sample). Although 2 of our targets show a ``rounder" morphology for the ENLR, this can be attributed to orientation effects, at least in the case of target 5, as discussed above. We thus argue that the preferred rounder morphologies for the ENLR found in previous  ground-based studies \citep[e.g.][]{liu13} may be due to the lack of angular resolution to resolve the biconical morphology. For example, in the case of target 2 (Fig.\,\ref{fig:panel2}), the region with low excitation perpendicular to the ionization axis, that defines the biconical morphology, has an angular width of only $\approx$\,0\farcs2, which is not resolved by previous ground-based studies without adaptative optics. The spatial resolution of the HST observations have been fundamental to reveal the presence of the torus at these high luminosities.

In order to further explore the ionization cones, we now discuss spatial one-dimensional line-ratio profiles obtained through the ENLR.

\subsubsection{[OIII]/H$\beta$ spatial profiles}
\label{o3hb}

Fig.\,\ref{fig:o3hb} shows spatial profiles of [OIII]/H$\beta$ passing through the nucleus along the ionization axis (blue and turquoise symbols) and perpendicularly to it (red and orange symbols) in logarithmic units. Also shown are the corresponding spatial profiles of the [OIII] fluxes along the ionization axis (black and grey lines). A trend can readily be seen for the line ratios: while along the ionization axis (blue symbols) the values are the highest, in the range 0.9\,$\le$log([OIII])/H$\beta\le$1.2, corresponding to 8\,$\le$[OIII]/H$\beta\le$16, along the perpendicular direction (the torus) they are lower, in the range 0.5\,$\le$log([OIII])/H$\beta\le$0.8, corresponding to 3\,$\le$[OIII]/H$\beta\le$7. 

Although we defer a more in-depth analysis of the excitation structure of the ENLR to future studies via resolved spectroscopy of our targets, preliminary results can be obtained by comparing the typical line ratios listed above for the ionization cone and the values perpendicular to it 
with those of photoionization models. We have used as reference the work of \citet{kv13}, and their photonization models constructed to reproduce the line ratios of the very extended emission nebula around the quasar MR\,2251-178, that shows an ionization cone extending to $\approx$\,90\,kpc from the nucleus. Adopting the [NII]/H$\alpha$ ratios of our targets from the SDSS spectra -- in the range 0.3--1, thus $-0.5\le$\,log([NII])/H$\alpha$)$\le$0 --  from Fig.\,3 of \citet{kv13}, panel (d) (dust-free photoionization models), for the typical range of the line ratios along the direction of the torus in our images, 
the ionization parameter value is log(U)$\approx-3$, while for the line ratios along the ionization cone, 
log(U)$\approx-2$. 

Considering that U$=$Q/(4$\pi$\,R$^2$\,n$_e$\,c$^2$), where Q is the rate of ionizing photons from the AGN, R is the distance from the nucleus and n$_e$ is the electronic density, and that for the inner region of the cones and the perpendicular direction n$_e$ and R should be similar, the difference in U is reflecting the difference in Q. This implies that the rate of ionizing photons escaping along the cones is $\approx$\,10 times the rate escaping perpendicularly to the cone, supporting the presence of an obscuring and collimating structure such as the torus.

Besides decreasing along the direction perpendicular to the ionization axis due to dilution of the radiation by the torus, the [OIII]/H$\beta$ ratios also decreases in certain regions due to other factors. In the case of target 3, there is a sharp decrease at $\approx$\,1\,kpc west of the nucleus -- to the right in Fig.\,\ref{fig:o3hb} -- in a region where the [OIII] flux is well above 3$\sigma_{sky}$. One possibility is the presence of a region of recent star formation. A similar effect -- the presence of regions of recent star formation, although smaller -- could explain small decreases in the [OIII]/H$\beta$ ratio along the ENLR in other objects.  Alternatively -- as we have estimated H$\beta$ from H$\alpha$, this decrease may be due to reddening, decreasing the [OIII] flux but less so the H$\alpha$ flux, leading to an overestimation of H$\beta$ relative to [OIII].

\subsubsection{The case for matter-bounded ENLR}

The [OIII]/H$\beta$ ratios decrease abruptely beyond the limits of the ENLR, where the [OIII] fluxes become lower than 3$\sigma_{sky}$, while the [OIII]/H$\beta$ ratios are remarkably constant througout the nebulae where the [OIII] is above 3$\sigma_{sky}$ as illustrated by the blue points in Fig. \,\ref{fig:o3hb}, suggesting a constant ionization parameter U. This can be understood if Q is constant and n$_e$ decreases with the distance from the nucleus as $R^{-2}$. Resolved measurements of the gas density along the ionization cone in the nearby Seyfert\,2 galaxy NGC\,3281 \citep{sb92} show exactly this behaviour for the gas density, supporting an approximately constant value of Q up to the limits of the nebulae, that we can define as corresponding to where the [OIII] flux decreases below 3$\sigma_{sky}$. This result suggests that the ENLR in all QSO 2s of this study are matter bounded.

The abrupt decrease in the [OIII]/H$\beta$ line ratios at the borders of the ENLR was also pointed out by \citet{liu13} in their GMOS-IFU data, arguing that this may imply that the ENLR is matter-bounded by the lack of gas at larger distances, including being bounded by the extent of the disk of the galaxy. This limit -- the galaxy radius -- should apply if there are no external sources of gas, 
but it could increase in the cases of availability of gas at larger distances, as seems to be the case in our sample.


Further evidence that the ENLR of the QSOs of the present study are matter bounded are present in at least three targets of our sample:
(1) in QSO 1 the ionizing radiation escapes the host galaxy and reaches a cloud to the south-east (Figs\,\ref{fig:panel1} and \ref{fig:o3hb}) ionizing the gas there. This implies that the apparent decrease in the gas excitation between the galaxy and the cloud is not due to the lack of photons, but to the lack of matter; (2) in QSO 2 the nuclear radiation ionizes gas up to $\sim10$\,kpc to the west (right in Figs.\,\ref{fig:panel2} and \ref{fig:o3hb}), by only to $\sim3$\,kpc to the east (left in the figures). This can be understood as the result of a merger process, during which gas is spread between the galaxies, and not as much to the other (east) side of the QSO, as seen in Fig. \ref {fig:panel2}. Thus, what is limiting the size of ENLR to the east in Fig.\,\ref{fig:o3hb} is not the rate of ionizing photons, but the availability of gas in that direction; (3) in QSO 4 the AGN radiation is ionizing a detached cloud to the right in Fig.\,\ref{fig:o3hb} (that could be the remnant of a merger, or the result of a previous outflow from the AGN, thus the decrease in excitation between the QSO and the cloud is due to the lack of gas, indicating that, also in the opposite direction, ionizing photons escape as the nebula is matter bounded.

The cases above indicate that the ionizing photons of the AGN can reach distances far beyond the apparent limits of the host galaxy as seen in the continuum images, what is supported also by previous studies in the literature such as the case of the quasar MR 2251-178 \citep{kv13} discussed above.


\subsection{The incidence of galaxy interactions}

The continuum images of the QSO hosts of our sample show signatures of interactions in 6 out of our 9 QSOs. In target 1 (Fig.\,\ref{fig:panel1}), there is a structure with angular size approximately 1$^{\prime\prime}\times1^{\prime\prime}$ in the continuum image at about\,15 kpc to the southeast of the AGN, but, due to the large contamination from gas emission in the continuum image of this galaxy (see Sec.\,\ref{sec:continuum}) we favor that this structure is a gas cloud only appering in the continuum image because of the [OIII] emission-line contribution. In target 2 (Fig.\,\ref{fig:panel2}), the QSO host is in clear interaction with a big companion closer than $\approx$10\,kpc to the southwest. In target 3  (Fig.\,\ref{fig:panel3}), the continuum image shows what seems to be a secondary nucleus $\approx$0\farcs3 from the nucleus to the southeast, suggesting this may be an advanced stage of a merger. Target 4 (Fig.\,\ref{fig:panel4}) does not show a clear signature of a merger, but there is a cloud to the northwest only seen in gas emission that could be either reminiscent of a merger/interaction or due to a previous outflow from the AGN. In target 6 (Fig.\,\ref{fig:panel6}), the continuum image shows a compact source at $\approx$ 20\,kpc to the northwest, with the QSO host showing an asymetric morphology elongated in this direction, that could be a result of an interaction. Target 8 (Fig.\,\ref{fig:panel8}) seems to be in a late stage of a merger with what seems to be remains of a galaxy $\approx$ 6\,kpc  to the northwest. Target 9 (Fig.\,\ref{fig:panel9}) also shows what seems to be a small galaxy at $\approx$ 5\,kpc to the south of the nucleus. 

In summary, 6 of the 9 QSOs show signatures of mergers or interactions. This frequency of interactions in the sample was not anticipated, as the QSOs were selected only for their proximity -- in order to allow resolving the NLR structure, and for their [OIII] luminosity -- in order to probe a higher luminosity regime than the previous spatially resolved studies. This high incidence of interactions support the relation between strong nuclear activity (above $10^{42.5}$\,erg\,s$^{-1}$) and galaxy mergers, as suggested by previous studies \citep[e.g.][]{treister12,glikman15}.

\subsection{Relation between R$_{maj}$ and L[OIII]}

The analysis of the [OIII] images from the HST NLR snapshot survey of \citet{sch03a, sch03b} revealed a relation between the NLR (or ENLR) extent R$_{NLR}$ and the AGN luminosity L([OIII]. For a sample of 60 AGN with L([OIII])$<\,10^{42}$ergs\,s$^{-1}$, they have found the  relation R$_{NLR}\propto$\,L[OIII]$^{0.33}$. This was  interpreted as due to the fact that the ionization parameter is not constant along the NLR (in which case the slope should be 0.5), and that most of the [OIII] emission comes from a low density region. 

A constant ionization parameter had been suggested by previous observations of the NLR of a sample of QSO's that implied a slope of $\sim$0.5 \citep{ben02}, later argued to be unphysical as would lead to predicted NLR sizes larger than the extent of the galaxy \citep{net04}. One possibility raised to explain the steeper slope obtained for the QSOs is a possible contamination by HII regions in the QSO host galaxy. More recent ground-based studies of the extent of the NLR in QSO2s \citep{gre11,liu13,hai13} have used compilations of NLR sizes obtained from ground-based imaging and long-slit spectroscopy, investigating also the slope of the relation between the NLR radius and its luminosity, finding a value $\alpha$=0.22. This has been interpreted as indicating that the NLR size is limited by the density and ionization state of the NLR gas rather than the availability of ionizing photons.

With our measurements we can now revisit the relation between the ENLR extent and the [OIII] luminosity. Combining our measurements of R$_{maj}$ with those of lower redshift and luminosity targets from \citet{ben02}, \citet{sch03b} and \citet{fischer18} we now span a much larger luminosity range -- 39$\le$log(L[OIII])$\le$43.3, L[OIII] in erg\,s$^{-1}$ -- from which we can derive a new relation. Although this relation has been recently investigated and expanded to higher luminosities by other authors \citep{gre11,liu13,hai13} as discussed above, these studies used ground-based heterogeneous data, that include long-slit spectroscopy and images obtained by different authors wiht different instruments. At the largest luminosities (and distances) of QSOs, the derived diameters in ground-based studies may be of the order of those of the PSFs, and thus carry large uncertainties.

\begin{figure}[htb!]
\includegraphics[width=1.\linewidth,angle=-0]{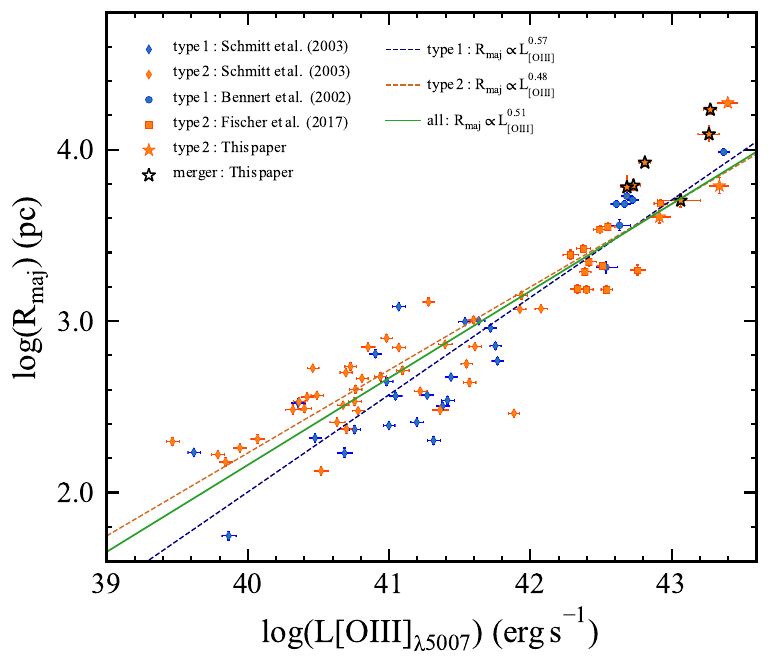}
\caption{Relation between the extent of the ENLR  R$_{maj}$ and log(L[OIII]), using data of three samples besides ours (stars): \citet{sch03b} (diamonds), \citet{ben02} (circles) and \citet{fischer18} (squares), distinguishing type 1 (blue) and type 2 (orange) AGN, together with the corresponding fits (see text). The AGN in our sample that appear to be in interaction are highlighted with black contours.}
\label{fig:logLxR}
\end{figure}

The higher luminosity data we have put together to derive a new relation were all obtained with HST, thus with at least 10 times better angular resolution than those from ground-based studies. This new relation is shown in Fig.\,\ref{fig:logLxR}, where the colors separate AGN type 1 (blue) and type 2 (orange), and our sample is represented as stars. The QSOs in our sample that appear to be in interaction are highlighted with black contours. 


Fig.\,\ref{fig:logLxR} shows that $R_{maj}$ continues to grow with $L[OIII]$ until the maximum luminosity of our sample -- $10^{43.3}\,erg\,s^{-1}$. Linear least square regressions were applied to the data using the  function \textit{curve\_fit} -- available in the Scipy library -- which also returns a covariance matrix whose diagonal value gives the standard deviation of the parameters fitted. We have obtained three regressions in order to verify also if we could detect any variation according to the AGN type. The first was restricted to type 1 AGN, returning the relation:
\begin{equation}
 \begin{aligned}
 \log(R_{maj}) ={} & (0.57\pm0.05)\log L[OIII]_{\lambda5007} \\
                 & -20.7\pm2.0,
 \end{aligned}
\end{equation}\label{eq:fitAGN1}
\noindent shown as a blue line in Fig.\,\ref{fig:logLxR}. The second was restricted to type 2 AGN:
\begin{equation}
 \begin{aligned}
 \log(R_{maj}) ={} & (0.48\pm0.03)\log L[OIII]_{\lambda5007} \\
                 & -17.1\pm1.1,
 \end{aligned}
 \end{equation}\label{eq:fitAGN2}
\noindent shown as an orange line in Fig.\,\ref{fig:logLxR}. And finally, using all AGNs we obtain:
\begin{equation}
 \begin{aligned}
 \log(R_{maj}) ={} & (0.51\pm0.03)\log L[OIII]_{\lambda5007} \\
                 & -18.12\pm0.98,
 \end{aligned}
 \end{equation}\label{eq:fitALL}
\noindent shown as a green line in Fig.\,\ref{fig:logLxR}. 

These regressions indicate that the dependency of the extent of the ENLR on L[OIII] is somewhat less steep in AGN type 2 than in AGN type 1, although, as there are only a few type 1 AGN with luminosities L[OIII]]$>10^{42.}$\,erg\,s$^{-1}$, this result is not very robust for high luminosities. The last fit, which uses all the data, returns a power of $0.51\pm0.03$, which is in agreement with the other two fits within the uncertainties, suggesting that there is no significant difference of this relation for the two types of AGN.

We have also investigated the effect in the relation of changing the measurements to exclude ``detached" gas clouds, that may be due to interactions. We considered, in particular, the cases of targets 1, 4, 8 and 9, for which, excluding the most external parts, $R_{maj} $ values change, respectively, from 18.8\,kpc to 9.2\,kpc, from 8.4\,kpc to 3.4\,kpc, from 17.1\,kpc to 14.0\,kpc and from 6.0\,kpc to 3.9\,kpc. A new fit, considering these new values result in:


\begin{equation}
 \begin{aligned}
 \log(R_{maj}) ={} & (0.49\pm0.03)\log L[OIII]_{\lambda5007} \\
                 & -17.4\pm0.94,
 \end{aligned}
\end{equation}\label{eq:fitAall_nc}

\noindent thus consistent with the previous fit within the uncertainties. We prefer to keep the previous fit because detailed imaging as we have in the present study is not always available, and in most cases of similar studies it will not be obvious if the extended gas emission is due to interactions or not.





The relation we have obtained for the whole sample is approximately the R$_{maj}\propto$L[OIII]$_{\lambda5007}^{1/2}$ previously obtained by \citet{ben02} for type 1 quasars at similar distances to those of our sources (0.1$\le$z$\le$0.4), supporting that the ENLR extent continues to grow with the AGN luminosity up to the highest AGN luminosity of our sample of  $\mathrm{L[OIII]=10^{43.3}\,erg\,s^{-1}}$.  This also seems to imply that the ionization parameter U may be constant, suggesting that the gas density $n_e$ decreases with distance from the nucleus (R) as $\sim$\,R$^{-2}$. This radial dependence for the gas density has been previously observed in NGC\,3281 \citep{sb92}, as already pointed out, and is close to the $\sim$\,R$^{-1.7}$ dependence also found for the NLR of NGC\,4151 \citep{kraemer00}, but is distinct for that observed in other cases in which the kinematics is clearly disturbed due to outflows \citep[e.g.][]{revalski18}. It may apply, though, to most of the ENLR, as the outflows seem to be restricted to the inner kpc or so \citep{fischer18}.

As pointed out above, \citet{net04} have argued for a limit in the size of the gaseous region ionized by the AGN corresponding to the extent of the galaxy, what makes sense if there is no gas beyond this limit. However, if there is gas spread beyond the galaxy as a consequence of outflows, interactions or mergers, the extent of the ENLR could still increase with the luminosity, provided that there are enough ionizing photons reaching these outer regions, what seems to be the case of our sample. Our results show that, as the AGN luminosity increases, the extent the ionized region continues to increase if there is gas to be ionized.

ENLRs extending even beyond those studied here have been reported in the literature, as is the case of the quasar MR 2251-178 \citep{kv13}, discussed in Sec.\,\ref{o3hb}, in which the ENLR reaches $\approx$\,90\,kpc from the nucleus. In addition, the authors conclude that the ENLR is still matter-bounded, as they estimate that ~65\%-95\% of the quasar ionizing radiation makes its way out of the system. As pointed out by these authors, this finding highlights the importance that quasar radiative feedback may have on the intergalactic medium and the need to for more similar studies.

In a long-slit spectroscopic study of a similar sample to ours, of 12 nearby (z$\sim\,0.1$) QSOs, \citet{sun17} report, for 7 galaxies of their sample, NLR extents ranging from 13 to 19\,kpc, overlaping our largest R$_{maj}$ values and reaching the same maximum value. But putting together their data with those from previous works of their group, they argue that the relation between the size of the NLR and the AGN luminosity flattens out at R$_{NLR}\sim10$\,kpc. We have not found this flattening in our data.

Alternatively, it may be argued that our sample is somehow ``special" due to the high occurrence of interactions, that provide gas to be ionized beyond the extent of the galaxies, even though they were selected solely on the basis of distance and luminosity. In addition, it is not obvious, at the corresponding galaxy distances or even larger, how to recognize if the gas origin is previous or on-going interactions.



\subsection{Total gas masses and surface mass densities}

The total masses of ionized gas are very similar among the galaxies, ranging from $0.4\times10^8$M$\odot$  to $2\times10^8$M$_\odot$. 
These masses are 2--3 orders of magnitude larger than those we have been obtaining in integral-field spectroscopic studies of nearby Seyfert galaxies (about 2 orders of magnitude less luminous) in the optical and near-infrared, although in these nearby cases the field-of-view correspond to smaller regions at the galaxies, of just a few kpcs \citep[e.g][]{rif15,rif18}. 

On the other hand, the ionized gas masses of our sample are similar to those obtained by \citet{har14} for a sample of 16 type 2 AGN of similar luminosities and redshifts to those of our targets and by \citet{tad14,couto17} for radio galaxies at similar redshifts, consistent with then having a similar origin, argued to be gas-rich mergers in \citet{tad14}.







\subsection{Implications for Feedback}
\label{feedback}

Our narrow-band images show extents for the ENLR ranging from 4\,kpc to 19\,kpc. These can be compared to those of \citet{fischer18}, for a similar sample of active galaxies also selected from \citet{rey08}, at somewhat lower redshifts and luminosities. \citet{fischer18} have also obtained HST [OIII] narrow-band images, as in our study, but instead of H$\alpha$+[NII] images, obtained long-slit STIS spectroscopy in order to investigate the [OIII] gas kinematics. The elongated morphologies of the ENLR are similar to those of our targets, but with somewhat lower R$_{maj}$ values, close to about 5\,kpc.

The long-slit spectroscopy of \citet{fischer18}, although obtained only along the ionization axis, revealed a disturbed kinematics for the gas suggesting outflows -- e.g. high velocity dispersion, double of triple components --  only within the inner kpc. Beyond this region, the gas kinematics, instead of outflows, revealed in most cases apparent rotation in the galaxy plane, with blueshifts observed to one side and redshifts to the other side of the nucleus. They measured the distance from the nucleus within which they have observed the signature of outflows, calling it R$_{out}$, finding that, on average, R$_{out}$/R$_{maj}=0.22$, where R$_{maj}$ is their adopted maximum extent of the ionized gas region (we have adopted the same notation as in their study). Typical outflow velocities are in the range 250--500 km\,s$^{-1}$.

The fact that the outflow is more compact than the ionized gas region in active galaxies is a common result also of previous studies of local AGN, such as those of the AGNIFS (AGN Integral Field Spectroscopy) group: \citet{schnorr14}, \citet{lena15}, \citet{rif15}, \citet{couto17}, as has also been pointed out by \citet{sun17}, \citet{fischer17} and \citet{revalski18}. This can be interpreted as due to the fact that most of the extended [OIII] emission comes from gas that is not outflowing,  but from gas that is usually rotating in the plane of the host galaxy. 

Although we do not have yet observed the gas kinematics of our sample, we can use the information we have gathered via the images and SDSS spectroscopy to estimate the mass outflow rate and the range of powers of the presumed outflows in our objects. In order to do this, we adopted the following assumptions.

(i) Regarding the extent of the outflow: our sample includes AGNs with higher luminosities than those of \citet{fischer18}, and it may well be that the outflow extends to larger distances than $R_{out}$. We thus considered two limits for the extents: an upper limit of R$_{maj}$ (which we called model $a$) and a lower limit of 0.2\,R$_{maj}$ (which we called model $b$). 

(ii) Regarding the outflowing gas mass: we have assumed that all gas mass within the considered radius ($R_{maj}$ or $R_{out}=0.2\,R_{maj}$) is outflowing.

(iii) Regarding the velocity of the outflow:  as the only spectroscopic data we have so far is the SDSS-III spectra, we have used the [OIII]$\lambda$5007 emission-line profile from these spectra to obtain the velocity $v_{80}=W_{80}/1.3$, where $W_{80}$ is the width of the line profile that comprises 80\% of the line flux, in the profile wings, thus probing the highest velocities, most probably due to the outflows, and the factor 1.3 is adopted to consider projection effects \citep{sun17}. This velocity is listed in the last column of Table\,\ref{tab:sdssfit}, and we have adopted it as the velocity of the outflow. In addition, we use the approximation that all spaxels have the same velocity, as we do not have spatial information on the kinematics so far.

We have used three methods to calculate the mass outflow-rate, and have used this rate to calculate also the power of the outflow in order to compare it with the QSO bolometric luminosity L$_{Bol}$. L$_{Bol}$ was calculated using the relation between the reddening-corrected L[OIII] and L$_{Bol}$ of \citet{trump15}. We have corrected L[OIII] for reddening according to equation 1 of \citet{lamastra09}, in which the average reddening value was obtained from the SDSS-III spectra H$\alpha$/H$\beta$ ratio assuming an intrinsic value of 3.0 \citep{ost06}. In order to compare the power of the outflows with L$_{Bol}$,  we have also corrected for reddening, in the calculations below, L(H$\beta$)  (assumed to have the same correction as [OIII] due to the proximity in wavelength between the two lines). These values were then used to obtain reddening-corrected values  for H$\alpha$ and M$_{ENLR}$.
\\

\noindent {\bf Method 1 --} The mass outflow rate was estimated from the ratio between the reddening-corrected mass of the ENLR $M$  and the time $t=R/v_{80}$ it took the gas at the border of the ENLR to reach the maximum distance R from the nucleus (R$_{maj}$ for model $a$ and 0.2\,R$_{maj}$ for model b). 

The power of the outflow  $dE/dt$  was calculated as:

\begin{equation}
\frac{dE}{dt}=0.5\,\frac{M\,v^2}{t}
\end{equation}\label{eq:power}

\noindent where the outflow velocity is assumed constant at $v=v_{80}$, and M is the value of M$_{ENLR}$ from Table\,\ref{tab:measurements} corrected for reddening.
\\





\noindent {\bf Method 2--} We use the same equation above to calculate the power of the outflow but, instead of calculating the mass-outflow rate as $M/t$, we use $(dM)/(dt)$ calculated as: 

\begin{equation}
\frac{dM}{dt}=m_p\,n_e\,v\,f\,A 
\end{equation}\label{eq:mass_rate1}

\noindent where $n_e$ is the electronic density assumed to be 100\,cm$^{-3}$ (as a typical value for extended emission-line regions, but see discussion below for the effect of a possibly higher density), $m_p$ is the mass of the proton, $f$ is the filling factor,  $v$ is the gas velocity, adopted as above ($v_{80}$), and $A$ is the cross-section of the outflow. 

We have then used the H$\alpha$ luminosity (obtained as $\eta$\,L([NII]+H$\alpha$), integrated within the radius considered, R$_{maj}$ or R$_{out}$ (for models $a$ and $b$, respectively) and corrected for reddening to derive $f$ :

\begin{equation}
L(H\alpha) = j_{H\alpha}\,4\pi\,f\,V \\
\end{equation}

\noindent where $j_{H\alpha}$ is the H$\alpha$ emissivity as defined in \citet{ost06}, and $V$ is the volume of the emitting region. We have first adopted as the geometry of the outflow a cone with base area $A$ and height $R$, assumed to be R$_{maj}$ in model $a$ and R$_{out}$ in model $b$ .  But we have nevertheless concluded that the geometry of the outflow cancels out when you calculate the filling factor, thus an outflow through a spherical surface with radius R would result in the same expression, as follows:

\begin{equation}
\dot{M}=\frac{3\,m_p \,v\,L(H\alpha)}{\left(\frac{j_{H\alpha}}{j_{H\beta}}\right) \,n_{e}\,\alpha_{H\beta}^{\rm eff}\,h\nu_{H\beta}\, R},
\end{equation}

\noindent with the emissivities $j$ and recombination coefficient $\alpha_{H\beta}^{eff}$ values obtained from \citet{ost06} for $n_e=100\,cm^{-3}$ and $T=10 000K$. We note that, in this method, the mass-outflow rate results inversely proportional to the gas density. Thus, if the the gas density is, for example, 5 times larger (500\,cm$^{-3}$), the mass outflow rate will be 5 times smaller. But there is no assuption regarding the filling factor, which is calculated from the data.
\\

\noindent {\bf Method 3 --} We have also estimated what can be considered a ``maximum" mass-outflow rate, using eq. 4 of  \citet{sun17}, that assumes a density, filling factor and a spherical geometry for the flow:

\begin{equation}
\frac{dM}{dt}= m_p\,n_e\,v\,4\,\pi\,R^2 
\end{equation}\label{eq:mass_rate3}

\noindent where  \citet{sun17}  adopt a constant $n_e=0.5$\,cm$^{-3}$, which is equivalent, for example, to the product of a density of $n_e=100$\,cm$^{-3}$ times a filling factor of 0.005 (a typical value, similar to what we have obtained from previous studies), or to a somewhat larger density of $n_e=500$\,cm$^{-3}$ times a smaller filling factor of 0.001.

The power of the outflow is obtained using again the equation\,\ref{eq:power} replacing $M/t$ by $dM/dt$ calculated as above.

The resulting mass outflow rates and powers are shown in Table\,\ref{tab:power} for the three methods above for each QSO. In the first line we show the power calculated assuming that the extent of the outflow is R$_{maj}$ and in the second line assuming it is R$_{out}$. We have also included in the table an estimate of the ratio between the outflow powers and the bolometric luminosity of the AGN $\dot{E}/L_{bol}$, where L$_{bol}$ was calculated from the L[OIII] value corrected for reddening  using the bolometric correction given by equation 9 of \citet{trump15}.

The values of the mass outflow rates $\dot{M}$ range from $\sim$5 to $\sim$200\,M$_\odot$\,yr$^{-1}$ for the first method and from $\sim$15 to $\sim$600\,M$_\odot$\,yr$^{-1}$ for the second method. A higher density of 500\,cm$^{-3}$ will result in a mass outflow rate 5 times lower, thus from $\sim$1 to $\sim$40\,M$_\odot$\,yr$^{-1}$ for the first method and $\sim$3 to $\sim$100\,M$_\odot$\,yr$^{-1}$ for the second method. These values range from being of the same order up to two orders of magnitude larger than those of our previous studies of closer and less luminous AGN \citep[e.g.][]{sto10,rif13,barbosa14}. In the work of \citet{revalski18}, in which the authors calculate the mass-outflow rate as a function of distance from the nucleus in the NLR of the Seyfert\,2 galaxy Mrk\,573, the highest value of $3.4\pm0.5$\,M$_\odot$\,yr$^{-1}$ is obtained at 210\,pc from the nucleus.

\begin{figure*}[htb!]
\centering
\includegraphics[width=.7\linewidth,angle=-0]{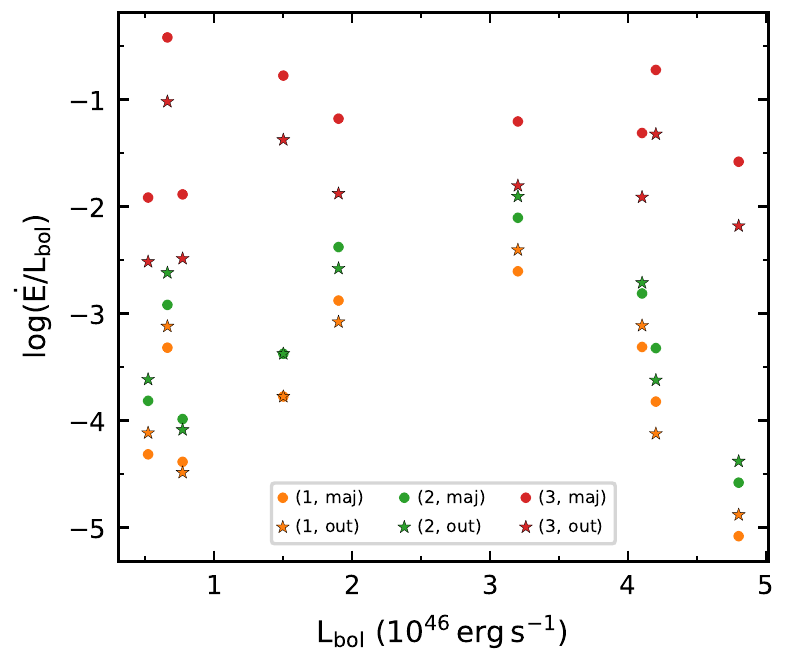}
\caption{Ratio between the outflow powers $\dot{E}$ and the AGN bolometric luminosity L$_{bol}$ as a function of L$_{bol}$ for methods 1 (yellow), 2  (green) and 3 (red) used to calculate $\dot{E}$ (see text). The filled circles correspond to an extent of the outflow of $R_{maj}$ while the star symbols correspond to an extent of $R_{out}=0.2\times R_{maj}$ (see text).}
\label{fig:power}
\end{figure*}

The corresponding estimated powers of the outflows (for methods 1 and 2) are in the range $10^{41.5} < $log$(\dot{E}) < $10$^{44.5}$ (in erg\,s$^{-1}$), and correspond to a  range of $ -5 < $log$(\dot{E}/L_{Bol}) < -2$, with only three cases in which it is larger than the 0.5\% threshold corresponding to a signficant impact on the host galaxy \citep[e.g.][]{hopelv10}. These $\dot{E}/L_{Bol}$ values are ploted as a function of L$_{Bol}$ in Fig.\,\ref{fig:power} in yellow and green symbols for methods 1 and 2, respectively. They increase between L$_{Bol}=0.5\times10^{46}$\,erg\,s$^{-1}$ and L$_{Bol}=3.5\times10^{46}$\,erg\,s$^{-1}$ but then decrease for higher luminosities.

It is important to point out that our calculations refer only to ionized gas, and do not consider the possible presence of neutral and molecular gas. Recent studies \citep[e.g.][]{fiore17} suggest that molecular gas is the dominant gas phase surrounding nearby AGN nuclei, with masses that can be $\sim$\,2 orders of magnitude larger than that of ionized gas. If such molecular gas is present in our sample and is also outflowing, the corresponding power of the outflow will be much larger than than the above and will most probably exceed the 0.5\% threshold. 

When using Method 3, in which the gas mass is not constrained by L(H$\alpha$) and the geometry is assumed to be spherical, the 
mass outflow rates reach values 2 to 3 orders or magnitude higher than those in Methods 1 and 2, corresponding thousands of solar masses per year. These mass outflow rates lead to estimated outflow powers that are above the 0.5\% threshold for most QSOs. The $\dot{E}/L_{Bol}$ values for Method 3 are ploted as a function of $L_{Bol}$ in Fig.\,\ref{fig:power} in red symbols. There seems to be no relation between these two quantities. As the mass outflow rate in this case is proportional to $R^2$, the resulting power is also proportional to $R^2$, and thus those calculated using $R_{maj}$ as the extent of the outflows are 5 times higher than those calculated using $R_{out}$, while this does not happen in the cases of Methods 1 and 2.

\begin{deluxetable*}{ccccccccccc}
\tablecolumns{11} 
\tablewidth{0pt} 
\tablecaption{Mass outflow rates and outflow power according to three different methods. \label{tab:power}}
\tablehead{
 &  & \multicolumn{3}{c}{Method 1}                                                                & \multicolumn{3}{c}{Method 2}                                                                & \multicolumn{3}{c}{Method 3}                                                                \\
\# & R & $\mathrm{ log(\dot{M})} $ & $\mathrm{log(\dot{E})} $ & $\mathrm{ log(\dot{E} / L_{bol})} $ & $\mathrm{ log(\dot{M})} $ & $\mathrm{ log(\dot{E})} $ & $\mathrm{ log(\dot{E} / L_{bol})} $ & $\mathrm{ log(\dot{M})} $ & $\mathrm{ log(\dot{E})} $ & $\mathrm{ log(\dot{E} / L_{bol})} $ }
\startdata 
1 & $\mathrm{R_{maj}}$ & 0.9 & 41.6 & -5.1 & 1.4 & 42.1 & -4.6 & 4.4 & 45.1 & -1.6 \\
  & $\mathrm{R_{out}}$ & 1.1 & 41.8 & -4.9 & 1.6 & 42.3 & -4.4 & 3.8 & 44.5 & -2.2 \\
2 & $\mathrm{R_{maj}}$ & 0.8 & 41.5 & -4.4 & 1.3 & 41.9 & -3.9 & 3.3 & 44.0 & -1.9 \\
  & $\mathrm{R_{out}}$ & 0.7 & 41.4 & -4.5 & 1.2 & 41.8 & -4.1 & 2.7 & 43.4 & -2.5 \\
3 & $\mathrm{R_{maj}}$ & 1.9 & 43.4 & -2.9 & 2.4 & 43.9 & -2.4 & 3.6 & 45.1 & -1.2 \\
  & $\mathrm{R_{out}}$ & 1.7 & 43.2 & -3.1 & 2.2 & 43.7 & -2.6 & 3.0 & 44.4 & -1.8 \\
4 & $\mathrm{R_{maj}}$ & 1.1 & 42.5 & -3.4 & 1.5 & 42.9 & -2.9 & 4.0 & 45.4 & -0.4 \\
  & $\mathrm{R_{out}}$ & 1.3 & 42.7 & -3.1 & 1.8 & 43.2 & -2.7 & 3.4 & 44.8 & -1.0 \\
5 & $\mathrm{R_{maj}}$ & 2.2 & 43.9 & -2.6 & 2.6 & 44.4 & -2.1 & 3.5 & 45.3 & -1.2 \\
  & $\mathrm{R_{out}}$ & 2.3 & 44.1 & -2.4 & 2.8 & 44.6 & -1.9 & 2.9 & 44.7 & -1.8 \\
6 & $\mathrm{R_{maj}}$ & 1.1 & 42.4 & -3.8 & 1.6 & 42.8 & -3.3 & 4.2 & 45.4 & -0.7 \\
  & $\mathrm{R_{out}}$ & 1.1 & 42.4 & -3.8 & 1.6 & 42.8 & -3.3 & 3.6 & 44.8 & -1.3 \\
7 & $\mathrm{R_{maj}}$ & 1.8 & 43.3 & -3.3 & 2.3 & 43.8 & -2.8 & 3.8 & 45.3 & -1.3 \\
  & $\mathrm{R_{out}}$ & 2.0 & 43.5 & -3.2 & 2.4 & 43.9 & -2.7 & 3.2 & 44.7 & -1.9 \\
8 & $\mathrm{R_{maj}}$ & 1.5 & 42.8 & -3.8 & 2.0 & 43.3 & -3.3 & 4.6 & 45.9 & -0.7 \\
  & $\mathrm{R_{out}}$ & 1.2 & 42.5 & -4.1 & 1.7 & 43.0 & -3.6 & 4.0 & 45.3 & -1.3 \\
9 & $\mathrm{R_{maj}}$ & 0.9 & 41.4 & -4.3 & 1.4 & 41.9 & -3.8 & 3.3 & 43.8 & -1.9 \\
  & $\mathrm{R_{out}}$ & 1.1 & 41.6 & -4.1 & 1.5 & 42.1 & -3.6 & 2.7 & 43.2 & -2.5 \\
\enddata 
\tablecomments{
(1): R is the adopted extent of the outflow, $\dot{M}$ is the mass outflow rate in units of $\mathrm{M_{\odot}\,yr^{-1}}$ and the
power of the outflow $\dot{E}$ is in units of $\mathrm{erg\,s^{-1}}$.}
\end{deluxetable*}

\section{Conclusions}\label{sec:conclusions}

We have presented HST continuum, [OIII]$\lambda5007$ and H$\alpha$+[NII] narrow-band images, as well as [OIII]/H$\beta$ excitation maps and have obtained the extent, morphology and masses of the ionized gas in the ENLR of 9 type 2 AGN classified as QSO2 due to their high luminosities ($\log$L[OIII$]>42.5$, L in ergs s$^{-1}$). Under the assumption that the ionized gas is outflowing, we have also estimated the corresponding mass outflow rates and powers.  Our main findings are:

\begin{itemize}

\item The ENLR is spatially resolved in all galaxies and its extent ranges from 4 to 19\,kpc; the largest extents may be related to the fact that 6 of the 9 galaxies seem to be in interaction, with gas emission extending well beyond the body of the galaxy seen in the continuum maps;

\item The ENLR morphology is clearly elongated and bipolar in 6 of the 9 galaxies, and excitation maps [OIII]/(H$\alpha$+[NII]) show a biconical structure, with the highest excitation gas along the ionization axis and decreasing with increasing angular distance from it; the exceptional HST angular resolution was key to reveal this structure;

\item  The high emission-line ratios obtained along the ionization axes ([OIII]/H$\beta\ge10$), and the low values obtained perpendicularly to it through the nucleus ([OIII]/H$\beta\le7$) imply that the rate of ionizing photons escaping along the ionization axis is $\approx$\,10 times higher than that escaping  perpendicularly to it;

\item The above result supports the presence of an obscuring torus that can thus survive the corresponding AGN luminosities, up to  log(L[OIII])=43.3 (L in ergs s$^{-1}$);

\item We have re-visited the relation between the extent of the ENLR (R$_{maj}$) and L[OIII], finding that R$_{maj}$ increases with L[OIII] 
over the range $39.5<$log[OIII]$<43.5$ (L in erg\,s$^{-1}$), as:  $\log$(R$_{maj}) = (0.51\pm0.03)\log($L[OIII]$)-18.12\pm0.98$;

\item We do not find a flattening in the above relation up to L[OIII]=43.3 (L in erg\,s$^{-1}$), implying that the ENLR in luminous AGN can extend to distances beyond the limit of the galaxy if there is availability of gas, e.g. from outflows or interactions, as seems to be the case of most galaxies of our sample; 

\item We attribute the flattening of the above relation reported in previous studies to the fact that the ENLR is matter bounded in most AGN hosts; the constancy of the [OIII]/H$\beta$ along the ENLR observed in our targets support that it is matter-bounded even in them, implying that the AGN radiation is still escaping to the intergalactic medium and do so in most luminous AGN;


\item The ionized gas masses of the ENLR range from 0.4 to $2\times10^8$\,M$_\odot$; much of it may have been acquired in an interaction, and thus seem to be more related to the feeding of the AGN than to its feedback, but this can only be concluded via resolved spectroscopy of the ENLR to probe its resolved kinematics what will be our next step following the present study;

\item Although we are still in the process of acquiring resolved spectroscopy of our targets, we have estimated the NLR outflow velocities from the widths W$_{80}$ of the [OIII] emission-line profiles in SDSS spectra, combined with the above ENLR masses and extents to estimate the mass outflow rates and power of the outflows using 3 different methods, as follows;

\item For the first two methods in which the ionized gas masses are constrained by the measured H$\alpha$ luminosity, we obtain mass outflow rates in ionized gas in the range $\sim$1--100\,M$_\odot$\,yr$^{-1}$; 

\item The corresponding powers of the outflows are larger than 0.5\%\,L$_{Bol}$ -- the usually adopted threshold for significant impact on the host galaxy -- only for 4 QSOs of the sample; on the other hand, we point out that our calculations do not take into account other gas phases, like neutral and molecular gas (that seems to be the dominant phase in many nearby AGN), that will lead to larger feedback powers, if present;

\item Using a third method in which the gas masses are not constrained by the H$\alpha$ luminosity, and the assumed outflow geometry is spherical, we obtain much larger mass-outflow rates, that reach thousands M$_\odot$\,yr$^{-1}$, with corresponding outflow powers exceeding 0.5\% L$_{Bol}$ for most QSOs.

\end{itemize}

We plan to revisit the above calculations in a forthcoming paper using resolved spectroscopy of our targets, from which we will obtain gas densities and temperatures and model the ionization structure of the ENLR. We are now in the process of acquiring such data.
\acknowledgments
\section*{ACKNOWLEDGEMENTS}

We thank the referee, Joss Bland-Hawthorn, for the numerous valuable suggestions that led to a much improved discussion of our results. This work has used data from the Hubble Space Telescope, which is operated by AURA for NASA, as well as from the Sloan Digital Sky Survey. Funding for SDSS-III has been provided by the Alfred P. Sloan Foundation, the Participating Institutions, the National Science Foundation, and the U.S. Department of Energy Office of Science. The SDSS-III web site is http://www.sdss3.org/. TSB acknowledges support from the Brazilian funding agencies CNPq, CAPES and FAPERGS. The submission of the HST proposal and the corresponding data were partially acquired in a visit by TSB to the Harvard/Smithsonian Center for Astrophysics. L.F. Longo Micchi was financed in part by Coordena\c c\~ao de Aperfei\c coamento de Pessoal de N\'ivel Superior - Brasil (CAPES) (grant 88888.040697/2013-00).



\begin{thebibliography}{}
\expandafter\ifx\csname natexlab\endcsname\relax\def\natexlab#1{#1}\fi
\providecommand{\url}[1]{\href{#1}{#1}}

\bibitem[{{Antonucci}(1993)}]{ant93}
{Antonucci}, R. 1993, \araa, 31, 473

\bibitem[{{Baldwin} {et~al.}(1981){Baldwin}, {Phillips}, \&
  {Terlevich}}]{bal81}
{Baldwin}, J.~A., {Phillips}, M.~M., \& {Terlevich}, R. 1981, \pasp, 93, 5

\bibitem[{{Barbosa} {et~al.}(2014){Barbosa}, {Storchi-Bergmann}, {McGregor},
  {Vale}, \& {Rogemar Riffel}}]{barbosa14}
{Barbosa}, F.~K.~B., {Storchi-Bergmann}, T., {McGregor}, P., {Vale}, T.~B., \&
  {Rogemar Riffel}, A. 2014, \mnras, 445, 2353

\bibitem[{{Bennert} {et~al.}(2002){Bennert}, {Falcke}, {Schulz}, {Wilson}, \&
  {Wills}}]{ben02}
{Bennert}, N., {Falcke}, H., {Schulz}, H., {Wilson}, A.~S., \& {Wills}, B.~J.
  2002, \apjl, 574, L105

\bibitem[{{Bland-Hawthorn} {et~al.}(2013){Bland-Hawthorn}, {Maloney},
  {Sutherland}, \& {Madsen}}]{bh13}
{Bland-Hawthorn}, J., {Maloney}, P.~R., {Sutherland}, R.~S., \& {Madsen}, G.~J.
  2013, \apj, 778, 58

\bibitem[{{Capetti} {et~al.}(1996){Capetti}, {Axon}, {Macchetto}, {Sparks}, \&
  {Boksenberg}}]{cap96}
{Capetti}, A., {Axon}, D.~J., {Macchetto}, F., {Sparks}, W.~B., \&
  {Boksenberg}, A. 1996, \apj, 469, 554

\bibitem[{{Ciotti} {et~al.}(2010){Ciotti}, {Ostriker}, \& {Proga}}]{cio10}
{Ciotti}, L., {Ostriker}, J.~P., \& {Proga}, D. 2010, \apj, 717, 708

\bibitem[{{Couto} {et~al.}(2017){Couto}, {Storchi-Bergmann}, \&
  {Schnorr-M{\"u}ller}}]{couto17}
{Couto}, G.~S., {Storchi-Bergmann}, T., \& {Schnorr-M{\"u}ller}, A. 2017,
  \mnras, 469, 1573

\bibitem[{{Crenshaw} {et~al.}(2010){Crenshaw}, {Kraemer}, {Schmitt},
  {Jaff{\'e}}, {Deo}, {Collins}, \& {Fischer}}]{cre10}
{Crenshaw}, D.~M., {Kraemer}, S.~B., {Schmitt}, H.~R., {et~al.} 2010, \aj, 139,
  871

\bibitem[{{Crenshaw} {et~al.}(2003){Crenshaw}, {Kraemer}, {Gabel}, {Kaastra},
  {Steenbrugge}, {Brinkman}, {Dunn}, {George}, {Liedahl}, {Paerels}, {Turner},
  \& {Yaqoob}}]{crenshaw03}
{Crenshaw}, D.~M., {Kraemer}, S.~B., {Gabel}, J.~R., {et~al.} 2003, \apj, 594,
  116

\bibitem[{{Elitzur} {et~al.}(2014){Elitzur}, {Ho}, \& {Trump}}]{elitzur14}
{Elitzur}, M., {Ho}, L.~C., \& {Trump}, J.~R. 2014, \mnras, 438, 3340

\bibitem[{{Elvis}(2000)}]{elv00}
{Elvis}, M. 2000, \apj, 545, 63

\bibitem[{{Fabian}(2012)}]{fab12}
{Fabian}, A.~C. 2012, \araa, 50, 455

\bibitem[{{Falcke} {et~al.}(1998){Falcke}, {Wilson}, \& {Simpson}}]{fal98}
{Falcke}, H., {Wilson}, A.~S., \& {Simpson}, C. 1998, \apj, 502, 199

\bibitem[{{Ferrarese} \& {Ford}(2005)}]{fer05}
{Ferrarese}, L., \& {Ford}, H. 2005, \ssr, 116, 523

\bibitem[{{Ferruit} {et~al.}(2000){Ferruit}, {Wilson}, \& {Mulchaey}}]{fer00}
{Ferruit}, P., {Wilson}, A.~S., \& {Mulchaey}, J. 2000, \apjs, 128, 139

\bibitem[{{Fiore} {et~al.}(2017){Fiore}, {Feruglio}, {Shankar}, {Bischetti},
  {Bongiorno}, {Brusa}, {Carniani}, {Cicone}, {Duras}, {Lamastra}, {Mainieri},
  {Marconi}, {Menci}, {Maiolino}, {Piconcelli}, {Vietri}, \&
  {Zappacosta}}]{fiore17}
{Fiore}, F., {Feruglio}, C., {Shankar}, F., {et~al.} 2017, \aap, 601, A143

\bibitem[{{Fischer} {et~al.}(2013){Fischer}, {Crenshaw}, {Kraemer}, \&
  {Schmitt}}]{fis13}
{Fischer}, T.~C., {Crenshaw}, D.~M., {Kraemer}, S.~B., \& {Schmitt}, H.~R.
  2013, \apjs, 209, 1

\bibitem[{{Fischer} {et~al.}(2017){Fischer}, {Machuca}, {Diniz}, {Crenshaw},
  {Kraemer}, {Riffel}, {Schmitt}, {Baron}, {Storchi-Bergmann}, {Straughn},
  {Revalski}, \& {Pope}}]{fischer17}
{Fischer}, T.~C., {Machuca}, C., {Diniz}, M.~R., {et~al.} 2017, \apj, 834, 30

\bibitem[{{Fischer} {et~al.}(2018){Fischer}, {Kraemer}, {Schmitt}, {Longo
  Micchi}, {Crenshaw}, {Revalski}, {Vestergaard}, {Elvis}, {Gaskell}, {Hamann},
  {Ho}, {Hutchings}, {Mushotzky}, {Netzer}, {Storchi-Bergmann}, {Straughn},
  {Turner}, \& {Ward}}]{fischer18}
{Fischer}, T.~C., {Kraemer}, S.~B., {Schmitt}, H.~R., {et~al.} 2018, \apj, 856,
  102

\bibitem[{{Fox} {et~al.}(2015){Fox}, {Bordoloi}, {Savage}, {Lockman},
  {Jenkins}, {Wakker}, {Bland-Hawthorn}, {Hernandez}, {Kim}, {Benjamin},
  {Bowen}, \& {Tumlinson}}]{fox15}
{Fox}, A.~J., {Bordoloi}, R., {Savage}, B.~D., {et~al.} 2015, \apjl, 799, L7

\bibitem[{{Glikman} {et~al.}(2015){Glikman}, {Simmons}, {Mailly}, {Schawinski},
  {Urry}, \& {Lacy}}]{glikman15}
{Glikman}, E., {Simmons}, B., {Mailly}, M., {et~al.} 2015, \apj, 806, 218

\bibitem[{{Greene} {et~al.}(2011){Greene}, {Zakamska}, {Ho}, \&
  {Barth}}]{gre11}
{Greene}, J.~E., {Zakamska}, N.~L., {Ho}, L.~C., \& {Barth}, A.~J. 2011, \apj,
  732, 9

\bibitem[{{Hainline} {et~al.}(2013){Hainline}, {Hickox}, {Greene}, {Myers}, \&
  {Zakamska}}]{hai13}
{Hainline}, K.~N., {Hickox}, R., {Greene}, J.~E., {Myers}, A.~D., \&
  {Zakamska}, N.~L. 2013, \apj, 774, 145

\bibitem[{{Haniff} {et~al.}(1988){Haniff}, {Wilson}, \& {Ward}}]{han88}
{Haniff}, C.~A., {Wilson}, A.~S., \& {Ward}, M.~J. 1988, \apj, 334, 104

\bibitem[{{Harrison} {et~al.}(2014){Harrison}, {Alexander}, {Mullaney}, \&
  {Swinbank}}]{har14}
{Harrison}, C.~M., {Alexander}, D.~M., {Mullaney}, J.~R., \& {Swinbank}, A.~M.
  2014, \mnras, 441, 3306

\bibitem[{{Hickox} {et~al.}(2014){Hickox}, {Mullaney}, {Alexander}, {Chen},
  {Civano}, {Goulding}, \& {Hainline}}]{hickox14}
{Hickox}, R.~C., {Mullaney}, J.~R., {Alexander}, D.~M., {et~al.} 2014, \apj,
  782, 9

\bibitem[{{Hopkins} \& {Elvis}(2010)}]{hopelv10}
{Hopkins}, P.~F., \& {Elvis}, M. 2010, \mnras, 401, 7

\bibitem[{{Hopkins} \& {Quataert}(2010)}]{hop10}
{Hopkins}, P.~F., \& {Quataert}, E. 2010, \mnras, 407, 1529

\bibitem[{{Kormendy} \& {Ho}(2013)}]{kor13}
{Kormendy}, J., \& {Ho}, L.~C. 2013, \araa, 51, 511

\bibitem[{{Kraemer} {et~al.}(2000){Kraemer}, {Crenshaw}, {Hutchings}, {Gull},
  {Kaiser}, {Nelson}, \& {Weistrop}}]{kraemer00}
{Kraemer}, S.~B., {Crenshaw}, D.~M., {Hutchings}, J.~B., {et~al.} 2000, \apj,
  531, 278

\bibitem[{{Kreimeyer} \& {Veilleux}(2013)}]{kv13}
{Kreimeyer}, K., \& {Veilleux}, S. 2013, \apjl, 772, L11

\bibitem[{{Lamastra} {et~al.}(2009){Lamastra}, {Bianchi}, {Matt}, {Perola},
  {Barcons}, \& {Carrera}}]{lamastra09}
{Lamastra}, A., {Bianchi}, S., {Matt}, G., {et~al.} 2009, \aap, 504, 73

\bibitem[{{Lena} {et~al.}(2015){Lena}, {Robinson}, {Storchi-Bergman},
  {Schnorr-M{\"u}ller}, {Seelig}, {Riffel}, {Nagar}, {Couto}, \&
  {Shadler}}]{lena15}
{Lena}, D., {Robinson}, A., {Storchi-Bergman}, T., {et~al.} 2015, \apj, 806, 84

\bibitem[{{Liu} {et~al.}(2013){Liu}, {Zakamska}, {Greene}, {Nesvadba}, \&
  {Liu}}]{liu13}
{Liu}, G., {Zakamska}, N.~L., {Greene}, J.~E., {Nesvadba}, N.~P.~H., \& {Liu},
  X. 2013, \mnras, 436, 2576

\bibitem[{{Netzer} {et~al.}(2004){Netzer}, {Shemmer}, {Maiolino}, {Oliva},
  {Croom}, {Corbett}, \& {di Fabrizio}}]{net04}
{Netzer}, H., {Shemmer}, O., {Maiolino}, R., {et~al.} 2004, \apj, 614, 558

\bibitem[{{Novak} {et~al.}(2011){Novak}, {Ostriker}, \& {Ciotti}}]{novak11}
{Novak}, G.~S., {Ostriker}, J.~P., \& {Ciotti}, L. 2011, \apj, 737, 26

\bibitem[{{Osterbrock} \& {Ferland}(2006)}]{ost06}
{Osterbrock}, D.~E., \& {Ferland}, G.~J. 2006, {Astrophysics of gaseous nebulae
  and active galactic nuclei}

\bibitem[{{Peterson}(1997)}]{pet97}
{Peterson}, B.~M. 1997, {An Introduction to Active Galactic Nuclei}

\bibitem[{{Pogge}(1988)}]{pog88}
{Pogge}, R.~W. 1988, \apj, 328, 519

\bibitem[{{Revalski} {et~al.}(2018){Revalski}, {Crenshaw}, {Kraemer},
  {Fischer}, {Schmitt}, \& {Machuca}}]{revalski18}
{Revalski}, M., {Crenshaw}, D.~M., {Kraemer}, S.~B., {et~al.} 2018, \apj, 856,
  46

\bibitem[{{Reyes} {et~al.}(2008){Reyes}, {Zakamska}, {Strauss}, {Green},
  {Krolik}, {Shen}, {Richards}, {Anderson}, \& {Schneider}}]{rey08}
{Reyes}, R., {Zakamska}, N.~L., {Strauss}, M.~A., {et~al.} 2008, \aj, 136, 2373

\bibitem[{{Riffel} {et~al.}(2015){Riffel}, {Storchi-Bergmann}, \&
  {Riffel}}]{rif15}
{Riffel}, R.~A., {Storchi-Bergmann}, T., \& {Riffel}, R. 2015, \mnras, 451,
  3587

\bibitem[{{Riffel} {et~al.}(2013){Riffel}, {Storchi-Bergmann}, \&
  {Winge}}]{rif13}
{Riffel}, R.~A., {Storchi-Bergmann}, T., \& {Winge}, C. 2013, \mnras, 430, 2249

\bibitem[{{Riffel} {et~al.}(2008){Riffel}, {Storchi-Bergmann}, {Winge},
  {McGregor}, {Beck}, \& {Schmitt}}]{rif08}
{Riffel}, R.~A., {Storchi-Bergmann}, T., {Winge}, C., {et~al.} 2008, \mnras,
  385, 1129

\bibitem[{{Riffel} {et~al.}(2018){Riffel}, {Storchi-Bergmann}, {Riffel},
  {Davies}, {Bianchin}, {Diniz}, {Sch{\"o}nell}, {Burtscher}, {Crenshaw},
  {Fischer}, {Dahmer-Hahn}, {Dametto}, \& {Rosario}}]{rif18}
{Riffel}, R.~A., {Storchi-Bergmann}, T., {Riffel}, R., {et~al.} 2018, \mnras,
  474, 1373

\bibitem[{{Schmitt} {et~al.}(2003{\natexlab{a}}){Schmitt}, {Donley},
  {Antonucci}, {Hutchings}, \& {Kinney}}]{sch03a}
{Schmitt}, H.~R., {Donley}, J.~L., {Antonucci}, R.~R.~J., {Hutchings}, J.~B.,
  \& {Kinney}, A.~L. 2003{\natexlab{a}}, \apjs, 148, 327

\bibitem[{{Schmitt} {et~al.}(2003{\natexlab{b}}){Schmitt}, {Donley},
  {Antonucci}, {Hutchings}, {Kinney}, \& {Pringle}}]{sch03b}
{Schmitt}, H.~R., {Donley}, J.~L., {Antonucci}, R.~R.~J., {et~al.}
  2003{\natexlab{b}}, \apj, 597, 768

\bibitem[{{Schnorr-M{\"u}ller}
  {et~al.}(2014{\natexlab{a}}){Schnorr-M{\"u}ller}, {Storchi-Bergmann},
  {Nagar}, \& {Ferrari}}]{schnorr14a}
{Schnorr-M{\"u}ller}, A., {Storchi-Bergmann}, T., {Nagar}, N.~M., \& {Ferrari},
  F. 2014{\natexlab{a}}, \mnras, 438, 3322

\bibitem[{{Schnorr-M{\"u}ller}
  {et~al.}(2014{\natexlab{b}}){Schnorr-M{\"u}ller}, {Storchi-Bergmann},
  {Nagar}, {Robinson}, {Lena}, {Riffel}, \& {Couto}}]{schnorr14}
{Schnorr-M{\"u}ller}, A., {Storchi-Bergmann}, T., {Nagar}, N.~M., {et~al.}
  2014{\natexlab{b}}, \mnras, 437, 1708

\bibitem[{{Sharp} \& {Bland-Hawthorn}(2010)}]{sb10}
{Sharp}, R.~G., \& {Bland-Hawthorn}, J. 2010, \apj, 711, 818

\bibitem[{{Storchi-Bergmann} {et~al.}(2010){Storchi-Bergmann}, {Lopes},
  {McGregor}, {Riffel}, {Beck}, \& {Martini}}]{sto10}
{Storchi-Bergmann}, T., {Lopes}, R.~D.~S., {McGregor}, P.~J., {et~al.} 2010,
  \mnras, 402, 819

\bibitem[{{Storchi-Bergmann} {et~al.}(1992){Storchi-Bergmann}, {Wilson}, \&
  {Baldwin}}]{sb92}
{Storchi-Bergmann}, T., {Wilson}, A.~S., \& {Baldwin}, J.~A. 1992, \apj, 396,
  45

\bibitem[{{Su} {et~al.}(2010){Su}, {Slatyer}, \& {Finkbeiner}}]{su10}
{Su}, M., {Slatyer}, T.~R., \& {Finkbeiner}, D.~P. 2010, \apj, 724, 1044

\bibitem[{{Sun} {et~al.}(2017){Sun}, {Greene}, \& {Zakamska}}]{sun17}
{Sun}, A.-L., {Greene}, J.~E., \& {Zakamska}, N.~L. 2017, \apj, 835, 222

\bibitem[{{Tadhunter} {et~al.}(2014){Tadhunter}, {Dicken}, {Morganti},
  {Konyves}, {Ysard}, {Nesvadba}, \& {Ramos Almeida}}]{tad14}
{Tadhunter}, C., {Dicken}, D., {Morganti}, R., {et~al.} 2014, \mnras, 445, L51

\bibitem[{{Treister} {et~al.}(2012){Treister}, {Schawinski}, {Urry}, \&
  {Simmons}}]{treister12}
{Treister}, E., {Schawinski}, K., {Urry}, C.~M., \& {Simmons}, B.~D. 2012,
  \apjl, 758, L39

\bibitem[{{Trump} {et~al.}(2015){Trump}, {Sun}, {Zeimann}, {Luck}, {Bridge},
  {Grier}, {Hagen}, {Juneau}, {Montero-Dorta}, {Rosario}, {Brandt},
  {Ciardullo}, \& {Schneider}}]{trump15}
{Trump}, J.~R., {Sun}, M., {Zeimann}, G.~R., {et~al.} 2015, \apj, 811, 26

\bibitem[{{Veilleux} \& {Osterbrock}(1987)}]{vo87}
{Veilleux}, S., \& {Osterbrock}, D.~E. 1987, \apjs, 63, 295

\bibitem[{{Wang} {et~al.}(2009){Wang}, {Fabbiano}, {Karovska}, {Elvis},
  {Risaliti}, {Zezas}, \& {Mundell}}]{wan09}
{Wang}, J., {Fabbiano}, G., {Karovska}, M., {et~al.} 2009, \apj, 704, 1195

\bibitem[{{Wilson} {et~al.}(1993){Wilson}, {Braatz}, {Heckman}, {Krolik}, \&
  {Miley}}]{wil93}
{Wilson}, A.~S., {Braatz}, J.~A., {Heckman}, T.~M., {Krolik}, J.~H., \&
  {Miley}, G.~K. 1993, \apjl, 419, L61

\bibitem[{{Wilson} \& {Tsvetanov}(1994)}]{wil94}
{Wilson}, A.~S., \& {Tsvetanov}, Z.~I. 1994, \aj, 107, 1227

\end{thebibliography}

\end{document}